\documentclass[reprint, amsmath, amssymb, aps, prf]{revtex4-2}

\usepackage[utf8]{inputenc}
\usepackage[T1]{fontenc}
\usepackage{graphicx}
\usepackage{bm}
\usepackage{xcolor}
\usepackage[colorlinks, citecolor=blue, filecolor=blue, linkcolor=blue, urlcolor=blue]{hyperref}
\usepackage[strings]{underscore}

\makeatletter
\providecommand{\href@noop}[2]{#2}
\makeatother

\usepackage[usenames,dvipsnames]{xcolor}
\hypersetup{
colorlinks,
citecolor=blue,
filecolor=blue,
linkcolor=blue,
urlcolor=blue}

\newcommand{\la}{\left\langle}
\newcommand{\ra}{\right\rangle}
\newcommand{\be}{\begin{equation}}
	\newcommand{\ee}{\end{equation}}
\newcommand{\bse}{\begin{subequations}}
	\newcommand{\ese}{\end{subequations}}
\newcommand{\bea}{\begin{eqnarray}}
	\newcommand{\eea}{\end{eqnarray}}
\newcommand{\ba}{\begin{array}}
	\newcommand{\ea}{\end{array}}

\begin{document}


\title{Scaling in Supersonic Turbulence: Energy Spectra and Fluxes using High-Fidelity Direct Numerical Simulations}

\author{Harshit Tiwari}%
\thanks{These authors contributed equally to this work.}
 \author{Dhananjay Singh}
\thanks{These authors contributed equally to this work.}
 \affiliation{ 
Department of Physics, Indian Institute of Technology Kanpur, Kanpur 208016, India
}

\author{Mahendra K. Verma}%
\affiliation{ 
Department of Physics, Indian Institute of Technology Kanpur, Kanpur 208016, India
}
\affiliation{ 
Kotak School of Sustainability, Indian Institute of Technology Kanpur, Kanpur 208016, India
}

\author{Rajesh Ranjan}%
 \thanks{Corresponding author}\email{rajeshr@iitk.ac.in}
\affiliation{ 
Department of Aerospace Engineering, Indian Institute of Technology Kanpur, Kanpur 208016, India
}

\date{\today}

\begin{abstract}
Supersonic turbulence plays a central role in astrophysical and high-speed engineering flows, yet the mechanisms governing its energy transfers remain poorly understood. In this work, we present high-resolution ($1024^3$) direct numerical simulations (DNS) of forced compressible turbulence across a wide range of turbulent Mach numbers, from low subsonic to high supersonic ($M_t = 0.2$--$3.0$). Using the GPU-accelerated solver \texttt{DHARA} with a seventh-order, low-dissipation Targeted Essentially Non-Oscillatory (TENO) scheme, we accurately resolve both fine-scale turbulent eddies and sharp shock fronts.  Our results reveal a fundamental change in the energy cascade in the supersonic regime. As $M_t$ increases, the rotational kinetic energy spectrum steepens from a Kolmogorov-like $k^{-5/3}$ scaling toward a Burgers-like $k^{-2}$ scaling. Conversely, the compressive energy spectrum becomes noticeably shallower, deviating from Burgers-like scaling. We show that these spectral modifications are driven by a dominant cross-scale transfer of energy from solenoidal to compressive modes within the inertial range, accompanied by non-negligible contributions from pressure dilatation. Furthermore, we demonstrate that the scaling laws for the root-mean-square compressive velocity ($U_C$) and the compressive energy flux ($\Pi_C$) closely mirror those of classical Burgers turbulence. We also examine how key energy transfer statistics vary with forcing and $M_t$. While the normalized solenoidal and compressive energy injection rates exhibit only weak dependence on $M_t$, they depend strongly on the ratio of rotational to compressive forcing amplitudes. Furthermore, as $M_t$ increases, the normalized rotational dissipation decreases, whereas both the normalized compressive dissipation and pressure dilatation increase. These findings elucidate the intermodal energy cascade mechanisms, advancing the understanding of energy transfers in supersonic turbulence.

\end{abstract}

\maketitle


\section{Introduction}\label{introduction}

Supersonic turbulence is of fundamental interest in astrophysics~\cite{Carroll:book:Astrophysics, Elmegreen:ARAA2004,Mac:RMP2004}, space physics, and high-speed aerodynamics~\cite{Von:JSR2003}. Prime examples include supernovae explosions, star formation in molecular clouds~\cite{Krumholz:APJ2005,Padoan:APJ2011}, inertial confinement fusion~\cite{Fujisawa:PJA2021}, and hypersonic propulsion~\cite{Ingenito:AIAA2010}. However, despite this significance, the mechanisms of energy transfer in supersonic flows remain poorly understood. Unlike incompressible turbulence, supersonic flows involve intricate interactions among shocks, acoustic modes, and nonlinear advection, leading to fundamentally distinct physical behaviors. Previous studies have also shown that supersonic turbulence ($M_t > 1$) differs markedly from the subsonic regime ($M_t < 1$)~\cite{Kritsuk:ApJ2007,Schmidt:AA2009,Federrath:MNRAS2013}. 

For incompressible turbulence, a well-established theoretical framework describes the velocity spectrum~\cite{Kolmogorov:DANS1941Dissipation,Kolmogorov:DANS1941Structure,Kraichnan:JFM1959,Dar:PD2001,Verma:PR2004,Lesieur:book:Turbulence}. A cornerstone of this theory is Kolmogorov’s spectrum, $E(k) = K_\mathrm{Ko} \, \epsilon_\mathrm{Ko}^{2/3} \, k^{-5/3}$, which characterizes the distribution of kinetic energy among wavenumbers $k$ in the inertial range, where $\epsilon_\mathrm{Ko}$ is the constant energy flux and viscous dissipation rate, and $K_\mathrm{Ko}$ is Kolmogorov’s constant~\cite{Kolmogorov:DANS1941Dissipation,Kolmogorov:DANS1941Structure,Lesieur:book:Turbulence}. Moreover, the energy cascade in incompressible turbulence is fundamentally driven by local interactions, meaning that energy transfer occurs primarily between neighboring wavenumber shells~\cite{Kraichnan:JFM1971_2D3D,Domaradzki:PF1990,Zhou:PF1993,Zhou:PF1993b,Verma:Pramana2005S2S}. 

The primary flow structures in incompressible turbulence consist of eddies and vortical tubes. In contrast, compressibility fundamentally modifies both the flow organization and the underlying dynamics~\cite{Kida:JSC1990, Lele:ARFM1994, Erlebacher:TCFD1990, John:JFM2021}, leading to the emergence of shocks and localized compression regions as dominant features. A key distinction from the incompressible regime is the presence of a dilatational (compressive) velocity component and associated pressure dilatation effects~\cite{Kida:JSC1990, Sarkar:PF1992, Miura:POF1995, Praturi:POF2019}. Consequently, the velocity field can be decomposed into solenoidal (divergence-free) and dilatational (curl-free) components, which exhibit distinct dynamics. Their relative contributions are primarily governed by the turbulent Mach number, defined as $M_t = U / C_s$, where $U$ is the root-mean-square velocity and $C_s$ is the average speed of sound ~\cite{Kida:JSC1990, Jagannathan:JFM2016, Sarkar:PF1992}.
In the nearly incompressible limit [$(\delta \rho)/\rho \rightarrow 0$], \citet{Zank:PF1991} demonstrated that both velocity and density fields follow the $k^{-5/3}$ spectrum. At subsonic Mach numbers, numerical simulations have shown that the rotational velocity component exhibit nearly Kolmogorov-like $k^{-5/3}$ spectrum, while the compressive component exhibits a steeper $k^{-2}$ scaling due to the presence of shocklets~\cite{Kida:JSC1990,Wang:PRL2013,Schmidt:PRE2019,Sakurai:PF2024}.

Supersonic turbulence is considerably more complex than its subsonic counterpart, and its scaling laws remain a subject of active debate despite extensive numerical investigations. \citet{Kritsuk:ApJ2007} reported that, for isothermal turbulence at $M_t \approx 6$, both solenoidal and compressive components exhibit a $k^{-2}$ velocity spectrum, while the kinetic energy spectrum follows a shallower $k^{-3/2}$ scaling. They further proposed a modified Kolmogorov framework based on the density-weighted velocity $\rho^{1/3} \mathbf{u} $, whose spectrum was found to be largely insensitive to variations in Mach number. It should be noted, however, that these results were obtained under isothermal and inviscid assumptions, neglecting viscous dissipation and compressive heating. Subsequent work by \citet{Schmidt:AA2009} reported similar trends, indicating that the influence of forcing is reduced when expressed in terms of $ \rho^{1/3} \mathbf{u} $. In contrast, \citet{Federrath:AA2010} demonstrated a strong sensitivity of the scaling behavior to the nature of turbulent forcing, showing that the spectra of $\mathbf{u}$, $ \rho^{1/2} \mathbf{u} $, and $ \rho^{1/3} \mathbf{u} $ all steepen under compressive driving, with the kinetic energy spectrum approaching a $k^{-2}$ scaling. Furthermore, \citet{Federrath:MNRAS2013} established consistency between these numerical observations and the exact relations derived by \citet{Galtier:PRL2011}, which predict a scaling of $ \rho^{1/3} \mathbf{u} \propto k^{-19/9} $ in the presence of strong compressive modes. These observations highlight that the scaling laws of supersonic turbulence remain unresolved and are strongly influenced by both the forcing mechanism and the degree of compressibility.

 Energy fluxes in supersonic turbulence are another aspect that is poorly understood. Note that incompressible flows have a single energy flux, which is constant in the inertial range. Compressible turbulence, unlike its incompressible counterpart, has multiple energy fluxes for its different velocity components and internal energy~\cite{Wang:PRL2013,Schmidt:PRE2019}. Using a coarse-graining framework, \citet{Aluie:PRL2011,Aluie:PhysD2013} showed that kinetic energy transfer in compressible turbulence can remain constant within an inertial range, where nonlinear interactions are dominant. This is possible as long as the pressure-dilatation co-spectrum diminishes faster than $k^{-1}$. \citet{Aluie:ApJL2012} provided the first direct evidence for this condition, demonstrating that kinetic and internal energy budgets statistically decouple beyond a transitional coversion scale. Furthermore, \citet{Zhao:PRF2018} numerically demonstrated that the Favre decomposition is unique in satisfying the inviscid criterion, which ensures that viscous effects remain negligible at these large scales. These findings were further supported by simulations of Euler equations in 2013 by \citet{Kritsuk:JFM2013}, which demonstrated that a Kolmogorov-like cascade can exist even in highly compressible flows.

\citet{Graham:ApJ2010} proposed a spectral framework to study energy transfer in compressible magnetohydrodynamic turbulence. They derived explicit transfer terms and separated advective and compressive contributions. Later, \citet{Schmidt:PRE2019} applied this idea to hydrodynamic turbulence using the density-weighted velocity $\mathbf{w} = \sqrt{\rho}\,\mathbf{u}$ in large-eddy simulations. Their study showed that the velocity spectra, scaling laws, and fluxes depend strongly on the nature of forcing. Earlier, \citet{Dar:PD2001} and \citet{Verma:PR2004} developed the mode-to-mode transfer formalism for incompressible turbulence, which was later extended to compressible flows by \citet{Singh:PRF2025}. This extension makes it possible to calculate fluxes separately for rotational and compressive motions and to directly quantify cross-mode energy transfers.

The above discussion highlights that the scaling laws and energy cascade for supersonic turbulence remain an open question. A contributing factor is the scarcity of studies that solve the full Navier–Stokes equations at supersonic turbulent Mach numbers within a direct numerical simulation (DNS) framework, resolving down to the dissipation scale with both high spatial resolution and high‐order numerical schemes.
Most such investigations to date have relied on the Euler equations or employed turbulence modeling. Among the notable numerical studies of supersonic turbulence--including those by~\citet{Kritsuk:ApJ2007},~\citet{Schmidt:AA2009} and~\citet{Schmidt:PRE2019}--the piecewise parabolic method (PPM)~\cite{Colella:JCP1984} has been used on uniform or adaptively refined grids to examine scaling laws and intermittency. While PPM is robust for shocks, its inherent numerical dissipation compromises its use in DNS of compressible turbulence by damping small-scale fluctuations and shocklets~\cite{Pirozzoli:ARFM2011, Fu:JCP2016}.

To address these limitations, we perform high-resolution ($1024^3$) DNS of forced compressible turbulence for a range of turbulent Mach numbers, $M_t \in \{0.2, 0.7, 1.0, 1.4,$ $1.8, 3.0\}$, using our in-house GPU-accelerated code \texttt{DHARA}~\cite{Tiwari:PNAS2025}. We employ a low-dissipation, seventh-order TENO scheme~\cite{Fu:JCP2016} to accurately resolve both fine-scale eddies and shock structures. Statistically stationary turbulence is maintained using stochastic Ornstein--Uhlenbeck (OU) forcing~\cite{Schmidt:AA2009}, which allows control over the solenoidal-to-compressive forcing ratio. The key contributions of this work are as follows:

\begin{itemize}
\item Simulations are performed for more than 40 eddy turnover times to ensure converged statistics. Changes in flow structures due to pressure dilatation and baroclinic torque at high turbulent Mach numbers are analyzed.
\item The behavior of rotational and compressive energy spectra is examined across a wide range of turbulent Mach numbers, from low subsonic to high supersonic regimes. The influence of Mach number on spectral cascades is investigated.
\item Energy fluxes are calculated using the framework developed by \citet{Singh:PRF2025}. These calculations are used to study the effects of compressibility in energy transfer processes, particularly in supersonic turbulence.
\item Cross-energy transfer between solenoidal and compressive modes is quantified using mode-to-mode transfer analysis. Furthermore, the conversion of kinetic energy into internal energy through pressure dilatation and viscous dissipation is evaluated.
\item Scaling for compressive rms velocity ($U_C$) and compressive energy flux  ($\Pi_C$) are compared with those predicted by Burgers turbulence.
\item By combining results from all six simulations, overall trends in normalized dissipation rates, peak normalized fluxes, and scaling exponents of various energy spectra are established. The role of forcing in sustaining compressible turbulence is also investigated. 
\end{itemize}

The paper is organized as follows. Section~\ref{sec:system} details the numerical methods and simulation parameters, including forcing. In \S\ref{sec:results}, we present our results, focusing on flow structures, energy spectra, and inter-scale fluxes. Finally, we summarize our conclusions in \S\ref{sec:conclusions}.

\section{Numerical approach} \label{sec:system}
Direct numerical simulations (DNS) of statistically steady, compressible turbulence at different turbulent Mach numbers are performed in a cubic periodic domain of size $(2\pi)^3$, discretized on a uniform collocation grid. In this section, we detail the governing equations, numerical methodology, and simulation parameters.

\subsection{Governing equations}
The governing equations are the compressible Navier-Stokes equations, solved in the nondimensional form. The equations can be represented using tensorial notations as given below~\cite{Kida:JSC1990}:
\begin{gather}
\frac{\partial \rho}{\partial t} + \frac{\partial}{\partial x_i}(\rho u_i)  = 0,\label{eq:continuity}  \\
\frac{\partial}{\partial t}(\rho u_i) + \frac{\partial}{\partial x_j}(\rho u_i u_j  + \delta_{ij} p - \tau_{ij})  = \rho F_{i},\label{eq:momentum} \\
\frac{\partial E_T}{\partial t} + \frac{\partial}{\partial x_i} \left( u_i(E_T + p) - q_i - u_j \tau_{ij} \right) = \rho u_i F_i, \label{eq:energy}
\end{gather}
where $\rho$, $\mathbf{u}$, $p$, $T$, and $\mathbf{F}$ denote the density, velocity, pressure, temperature, and external force field, respectively. The viscous stress tensor $\tau_{ij}$ is defined as
\begin{equation}
\tau_{ij} = \frac{1}{\mathrm{Re}_0} \left( \partial_j u_i + \partial_i u_j - \frac{2}{3} \delta_{ij} \partial_m u_m \right),
\end{equation}
where $\mathrm{Re}_0$ is the reference Reynolds number.
The heat flux $q_i$ is given by Fourier's law:
\be
    q_i = \frac{1}{M_0^2 \mathrm{Pr} \mathrm{Re}_0 (\gamma -1)} \frac{\partial T}{\partial x_i}.
\ee
The total energy density $E_T$ is composed of the kinetic energy density $E_u$ and internal energy density $I$:
\be
E_T = E_u + I, \quad \text{with} \quad E_u = \frac{\rho u^2}{2}, \quad I = \frac{p}{\gamma - 1}.
\ee

\subsubsection{Non-dimensionalization}
The equations have been nondimensionalized using reference quantities: density $\rho_0$, temperature $T_0$, velocity $u_0$, and length $l_0$. The dimensionless numbers, thus, governing  the system are:
\begin{gather}
    \mathrm{Reynolds~number}~~\mathrm{Re}_0 = \frac{\rho_0 u_0 l_0}{\mu} , \\
    \mathrm{Mach~number}~~M_0 = \frac{u_0}{c} = \frac{u_0}{\sqrt{\gamma R^* T_0}}, \\
    \mathrm{Prandtl~number}~~\mathrm{Pr} = \frac{\mu C_p}{K_c}
\end{gather}
where $c$ is the speed of sound, $R^*$ is the specific gas constant, $\mu$ is the dynamic viscosity,and $K_c$ is the thermal conductivity~\cite{Kida:JSC1990,Jagannathan:JFM2016}. Additionally, we define two more non-dimensional numbers that characterize the system,

\begin{gather}
\mathrm{Taylor~Reynolds~number}~~\mathrm{Re}_\lambda = \bigg(\frac{5}{3 \mu \epsilon } \bigg)^{1/2}  \rho_0 U^2, \\
    \mathrm{Turbulent~Mach~number}~~M_t = \frac{U}{ c}
\end{gather}
 where $\epsilon$ is the mean viscous dissipation rate and $U$ is the root-mean-square velocity. The average dissipation rate is calculated as
\be
\epsilon = -\la u_j \frac{\partial \tau_{ij}}{\partial x_j} \ra.
\ee 
Based on the non-dimensionalization, the ideal gas law can be given as:
\begin{equation}
p = \frac{\rho T}{\gamma M_0^2},
\end{equation}
where $\gamma = C_p/C_v$ is the ratio of specific heats.

\subsubsection{External forcing} \label{subsec:forcing}
We apply an external stochastic forcing at large scales to sustain turbulence, following a generalised Ornstein–Uhlenbeck (OU) process in Fourier space similar to \citet{Schmidt:AA2009}. The evolution of the Fourier-transformed force $\mathbf{\hat{F}}(\mathbf{k},t)$ is governed by the stochastic differential equation:
\be
    \mathrm{d}\mathbf{\hat{F}}(\mathbf{k},t) = g_\zeta \Bigg[ -\mathbf{\hat{F}}(\mathbf{k},t)\frac{\mathrm{d}t}{t_0}
    +  \mathbf{P}_\zeta(\mathbf{k}) \cdot  [f_0(\mathbf{k}) \mathrm{d}\mathbf{\mathcal{W}}_t] \Bigg].
\ee
The first term on the right-hand side represents a linear damping with a relaxation timescale $t_0$, which ensures the forcing is correlated in time. The second term is the stochastic driving, where $\mathrm{d}\mathbf{\mathcal{W}}_t$ is a vector-valued Gaussian random variable with zero mean and variance $\mathrm{d}t$. The term $f_0(\mathbf{k})$ is the forcing amplitude, defined as 
\be
f_0(\mathbf{k}) = 2U_0^2 \left( \frac{2U_0 \sigma^2(\mathbf{k})}{\sqrt{\pi/2}} \right)^{1/2},
\ee
where $U_0$ is a characteristic velocity scale. The spectral profile of the forcing is determined by $\sigma(k)$, given as
\be
\sigma(k) \propto (k - k_{\min})^2 (k_{\max} - k)^2,
\ee
for $k_{\min} \leq k \leq k_{\max}$ and zero elsewhere.  

The projection operator $\mathbf{P}_\zeta(\mathbf{k})$ projects the stochastic vector $f_0(k) \mathrm{d}\mathbf{\mathcal{W}}_t$ to control the solenoidal and compressive components of the forcing. Specifically, $\zeta = 1$ corresponds to purely solenoidal forcing (divergence-free), whereas $\zeta = 0$ yields purely compressive forcing (curl-free). In tensor notation, the operator is defined as~\cite{Schmidt:AA2009}
\bea
(P_{ij})_\zeta(\mathbf{k}) &=& \zeta P_{ij}^\perp(\mathbf{k}) + (1 - \zeta) P_{ij}^{\parallel}(\mathbf{k}) \nonumber \\
&& = \zeta \delta_{ij} + (1 - 2\zeta) \frac{k_i k_j}{k^2},
\eea
where $\delta_{ij}$ is the Kronecker symbol, and $P_{ij}^\perp$ and $P_{ij}^{\parallel}$ are the fully solenoidal and the compressive projection operators, respectively. The normalization factor $ g_\zeta $ ensures that the energy injection rate remains independent of the solenoidal-compressible mix set by $ \zeta $. It is given by
\be
g_\zeta = \frac{1}{\sqrt{1 - 2\zeta + 3\zeta^2}}.
\ee
In our simulations, we choose $k_{\min} = 1$ and $k_{\max} = 3$, which results in a peak forcing amplitude at $k_0 = 2$. The integral length scale of the forcing is approximately $\ell \sim \pi$, i.e., half the domain size $2\pi$.

\subsection{Numerical method}\label{sec:method}
We solve Eqs.~(\ref{eq:continuity}-\ref{eq:energy}) using finite volume method on a structured uniform Cartesian grids. All the equations are written in vectorial notation in the following conservative form~\cite{Anderson:book:CFD}:
\be
    \frac{\partial \mathbf{X}}{\partial t} + \sum_{\beta} \frac{\partial \mathbf{F}^\beta}{\partial x_\beta}
    = \sum_{\beta} \frac{\partial \mathbf{G}^\beta}{\partial x_\beta} + \mathbf{S}, \label{eq:conservation}
\ee
where $\mathbf{X}$ is a column vector that contains the variables $\rho$, $\rho u_\beta$, and $E_T$, and the index $\beta = 1,2,3$ represent the $x$, $y$, and $z$ directions respectively. Here, $\mathbf{F}^\beta$ represents the nonlinear convection flux in the $\beta$-direction, $\mathbf{G}^\beta$ is the corresponding viscous flux, and $\mathbf{S}$ is the source term. Hence,
\bea
    \mathbf{X} &=
    \begin{bmatrix}
    \rho \\ \rho u_x \\ \rho u_y \\ \rho u_z \\ E_T
    \end{bmatrix}, \quad
    \mathbf{F}^\beta =
    \begin{bmatrix}
    \rho u_\beta \\
    \rho u_\beta u_x + \delta_{\beta x} p \\
    \rho u_\beta u_y + \delta_{\beta y} p \\
    \rho u_\beta u_z + \delta_{\beta z} p \\
    u_\beta (E_T + p)
    \end{bmatrix}, \nonumber  \\
    \mathbf{G}^\beta &=
    \begin{bmatrix}
    0 \\
    \tau_{\beta x} \\
    \tau_{\beta y} \\
    \tau_{\beta z} \\
    \sum_{\beta} u_\beta \tau_{\beta \beta} + q_\beta
    \end{bmatrix}, \quad
    \mathbf{S} =
    \begin{bmatrix}
    0 \\
    \rho F_x \\
    \rho F_y \\
    \rho F_z \\
    \sum_{\beta} u_\beta F_\beta
    \end{bmatrix}.
\eea

The convective fluxes are evaluated using the semidiscrete central scheme of Kurganov and Tadmor~\cite{Kurganov:JCP2000, Kurganov:JSC2000}, which avoids the need for Riemann solvers while retaining robustness near discontinuities. In the $x$-direction, the reconstructed left ($q^-_{i+1/2,j,k}$) and right ($q^+_{i+1/2,j,k}$) states are used to compute the numerical flux,
\bea
    \mathbf{F}^x_{i+1/2,j,k} &=& \frac{1}{2} \left[ \mathbf{F}^x(q^-_{i+1/2,j,k}) + \mathbf{F}^x(q^+_{i+1/2,j,k}) \right] - \nonumber \\
    && \frac{a_{i+1/2,j,k}}{2} \left( q^+_{i+1/2,j,k} - q^-_{i+1/2,j,k} \right),
\eea
where $a_{i+1/2,j,k}$ is the local maximum propagation speed in $x$-direction determined from the eigenvalues of the Jacobian flux. Identical processes are also adopted for the $y$- and $z$-directions. The viscous fluxes $\mathbf{G}^\beta$ are discretized using fourth-order central difference schemes. Time integration is performed using the explicit third-order Strong Stability Preserving Runge-Kutta (eSSPRK3) method~\cite{Gottlieb:SIAM2001}. The timestep $\Delta t$ is dynamically adjusted according to the Courant–Friedrichs–Lewy (CFL) condition,
\begin{equation}
    \Delta t = C_0 \min_{\beta} \left( \frac{\Delta x_\beta}{\max |\Lambda_\beta|} \right),
\end{equation}
where $C_0$ is CFL number and $\Lambda_{\beta}$ are the spectral radii of the Jacobian flux in the $\beta$ directions.

In the seminal work of~\citet{Kurganov:JCP2000}, a linear reconstruction of the cell-average was used to compute left and right interface states. To achieve high-order spatial accuracy, we tested several nonlinear reconstruction schemes that are widely used in the literature, including \textit{Weighted Essentially Non-Oscillatory} (WENO)~\cite{Jiang:JCP1996, Borges:JCP2008}, \textit{Central WENO} (CWENO)~\cite{Levy:SJSC2000,Kurganov:JSC2000} and \textit{Targeted Essentially Non-Oscillatory} (TENO)~\cite{Fu:JCP2016} schemes. Classical WENO formulations, such as WENO-JS \cite{Jiang:JCP1996}, can be overly dissipative in smooth flow regions. To mitigate this, we initially considered WENO-Z \cite{Borges:JCP2008}, which utilizes global smoothness indicators to reduce dissipation. However, we ultimately adopted the TENO scheme, which replaces the weighted stencil averaging of WENO with a threshold-based cutoff strategy to sharply discard non-smooth stencils~\cite{Fu:JCP2016}. The numerical methods described are implemented using the high-performance Python fluid solver \texttt{DHARA} \cite{Tiwari:PNAS2025}. In Appendix ~\ref{sec:dhara}, benchmark tests for various high-order reconstruction schemes are briefly presented for 3D compressible Taylor-Green Vortex (\S ~\ref{sec:tgv}), Isentropic Vortex (\S ~\ref{sec:isev}), and Kelvin-Helmholtz instability (\S ~\ref{sec:khi}). For all these cases, we found the seventh-order Targeted Essentially Non-Oscillatory (TENO7) scheme to be the most accurate, exhibiting the lowest dissipation while maintaining robustness. 

Designed for modern supercomputers, \texttt{DHARA} leverages $\texttt{CuPy}$ \cite{cupy_learningsys2017} for seamless GPU acceleration via NVIDIA CUDA, achieving a speedup of around $150X$ on a single NVIDIA A100 GPU compared to a single-core AMD EPYC 7543 CPU. The solver demonstrates excellent scalability on systems like Frontier (Oak Ridge National Laboratory) and Polaris (Argonne National Laboratory), showing strong scaling and good weak scaling across multiple nodes (see Appendix~\ref{sec:dhara}).

\subsection{Simulation details}
The simulations are performed at six turbulent Mach numbers, $M_t = 0.2, 0.7, 1.0, 1.4, 1.8$ and $3.0$. To sustain turbulence, we apply external stochastic forcing following an Ornstein–Uhlenbeck process, as detailed in \S~\ref{subsec:forcing}. We inject energy in the wavenumber range $1 \le k \le 3$, with the peak amplitude at $k_0 = 2$. The forcing is governed by two parameters: the projection parameter $\zeta$, which determines the degree of compressibility, and the amplitude $U_0$, which controls the root-mean-square velocity. For the simulations with $M_t = 0.2, 1.0, 1.8,$ and $3.0$, we employ predominantly solenoidal driving ($\zeta = 2/3$) with amplitudes $U_0 =0.02$, $0.1$, $0.2$ and $0.3$ respectively. Conversely, for the $M_t = 0.7$ and $1.4$ runs, we use a more compressively weighted forcing ($\zeta = 1/3$) with $U_0 = 0.1$ and $0.2$, respectively. These parameters were selected to study steady-state turbulence across a range of $M_t$ and investigate the effects of varying compressibility in the external driving.






 We choose $\gamma = 1.00001$, making the flow nearly isothermal. This ensures a statistically stationary state by preventing viscous dissipation from monotonically increasing the internal energy and reducing the Mach number. In addition, the nondimensionalized gas constant $R^* = 1/(\gamma M_0^2) \approx 0.01$ leads to $\sigma \approx 0.01 \rho T$ and $I = \sigma/(\gamma-1) \approx 10^3$. This models astrophysical flows with efficient radiative cooling, such as supersonic turbulence in the interstellar medium and molecular clouds. The other simulation parameters are set as $Pr = 1$ and $M_0 = 10$. In each simulation, the flow is initialized with uniform density and temperature, $\rho(t=0) = T(t=0) = 1$, and zero velocity $\mathbf{u}(t=0) = 0$. The simulations are evolved for more than 40 eddy turnover times, $t_{\text{eddy}} = l_0/u_0$, to ensure statistical convergence.  To speed up convergence and reduce initial transients, we first evolve the system on a coarser $512^3$ grid until it reaches a statistically steady state. The resulting field is then interpolated onto a finer $1024^3$ grid for further evolution. Each high-resolution run was performed on the Polaris supercomputer with 32 nodes (128 A100 GPUs). The highest Mach number run $M_t=3.0$, required approximately 60 hours of computation time.

 \begin{table}
\centering
\footnotesize
\setlength{\tabcolsep}{3pt}
\renewcommand{\arraystretch}{1.4}
\caption{Simulation parameters for all runs. Listed are the forcing parameter $\zeta$, characteristic velocity scale $U_0$, energy injection rates $(\epsilon^{\mathrm{inj}}_R, \epsilon^{\mathrm{inj}}_C)$, Taylor microscale Reynolds number ($\mathrm{Re}_\lambda$), turbulent Mach number ($M_t$), resolution parameter ($\eta/\Delta x$), and total time of simulation $t_{\mathrm{run}}$ in eddy turnover time.}
\vspace{4pt}
\begin{tabular}{ccccccccc}
\hline
Run & $\zeta$    & $U_0$  & $\epsilon^{\mathrm{inj}}_R$  & $\epsilon^{\mathrm{inj}}_C$ & $\mathrm{Re}_\lambda$ & $M_t$ & $\eta/\Delta x$ & $t_{\mathrm{run}}$ \\
\hline
1 &  $2/3$ &  $0.02$ & $4.51\times10^{-7}$ & $8.82\times10^{-9}$ &  $251$ & 0.2 & 1.16 & 44 \\

2 & $1/3$ & $0.1$ & $3.87\times10^{-5}$ & $1.24\times10^{-5}$ &  $245$ & 0.7 & 1.14 & 46 \\

3 & $2/3$ & $0.1$ & $1.02\times10^{-4}$ & $7.76\times10^{-6}$ &  $226$ & 1.0 & 1.13 & 44 \\

4 &  $1/3$ &  $0.2$ & $3.08\times10^{-4}$ & $0.92\times10^{-4}$ &  $181$ & 1.4 & 1.26 & 84 \\

5 & $2/3$ & $0.2$ & $1.13\times10^{-3}$ & $6.16\times10^{-5}$ &  $195$ & 1.8 & 1.27 & 84 \\

6 & $2/3$ & $0.3$ & $4.23\times10^{-3}$ & $1.31\times10^{-4}$ &  $220$ & 3.0 & 1.32 & 44 \\
\hline
\end{tabular}
\label{table:parameters}
\end{table}
 
Table~\ref{table:parameters} summarizes the key parameters for the these simulations, including the forcing parameter $\zeta$, characteristic velocity scale $U_0$, energy injection rates $(\epsilon^{\mathrm{inj}}_R, \epsilon^{\mathrm{inj}}_C)$, Taylor microscale Reynolds number $\mathrm{Re}_\lambda$, steady-state turbulent Mach number $M_t$, and the Kolmogorov scale resolution ratio $\eta/\Delta x$. Here, $\eta = \left[{\langle \mu \rangle^3}/({\epsilon \langle \rho \rangle^2)} \right]^{1/4}$
is the Kolmogorov length scale, with $\langle \mu \rangle$ and $\langle \rho \rangle$ denoting volume-averaged dynamic viscosity and density, and $\epsilon$ the average energy dissipation rate~\cite{Jagannathan:JFM2016}. We verify that all runs are well-resolved, satisfying the standard resolution criterion $\eta/\Delta x \geq 1$ (see Table~\ref{table:parameters}).

\section{Results and discussion}\label{sec:results}
\label{sec:num_results}
This section reports our numerical results, including global quantities, flow structures, energy spectra, and energy fluxes. We start with the global quantities and flow structures.



\subsection{Global quantities and flow structures}

\begin{figure*}
    \centering
    \includegraphics[width=0.9\textwidth]{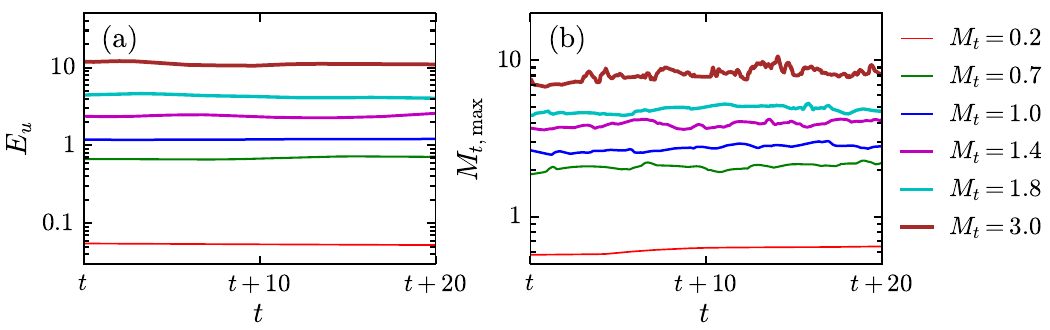}
    \caption{Time evolution of (a) total kinetic energy $E_u$ and (b) maximum Mach number $M_{t,\mathrm{max}}$ in the statistically stationary  state for runs with $M_t = 0.2$ (red), $0.7$ (green), $1.0$ (blue), $1.4$ (magenta), $1.8$ (cyan), and $3.0$ (brown).}\label{fig:steady}
\end{figure*}

\begin{figure*}
    \centering
    \includegraphics[width=1\textwidth]{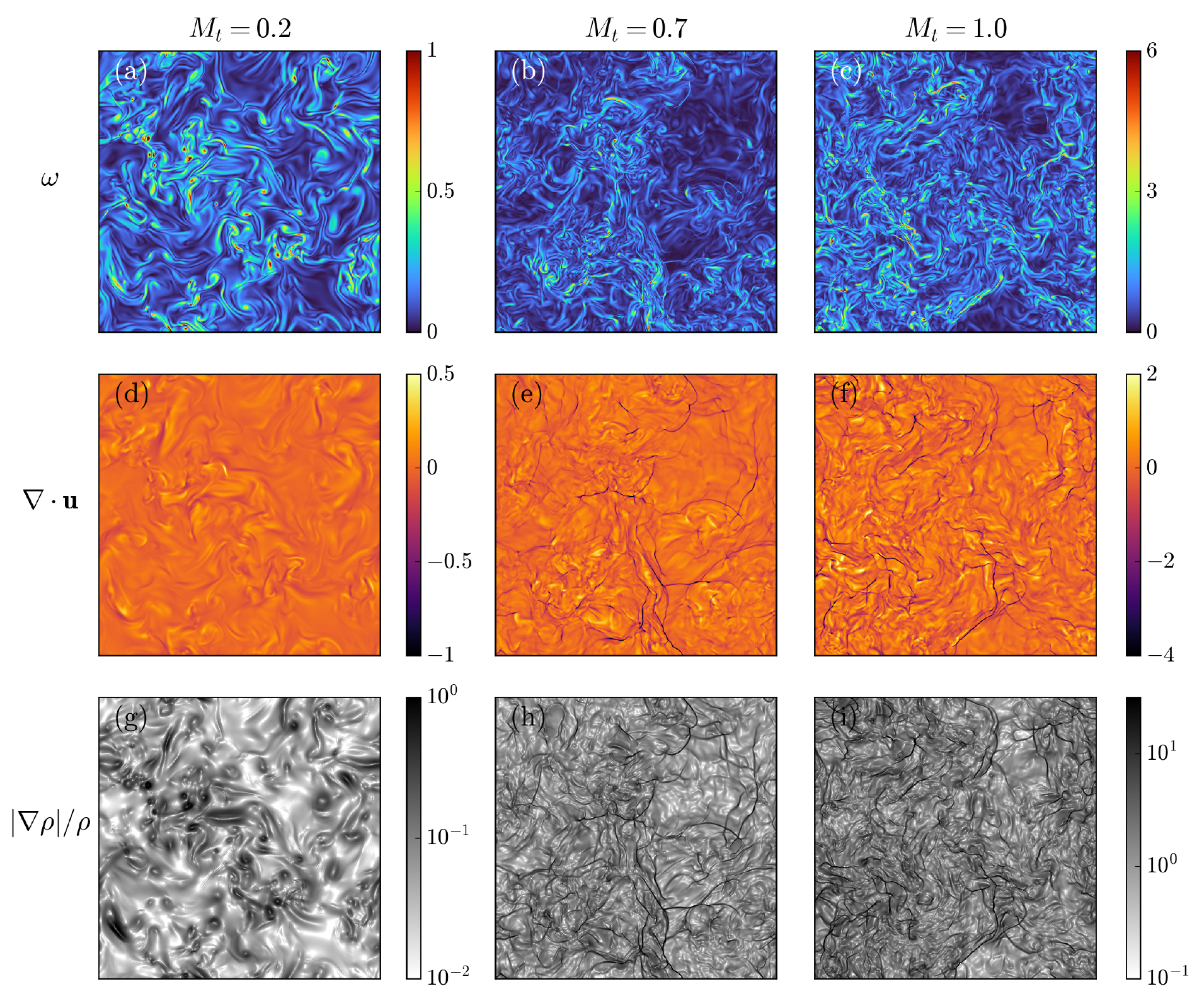}
    \caption{Flow structures in subsonic and transonic turbulence at $M_t = 0.2$ (left), $0.7$ (middle), and $1.0$ (right), shown on $x$–$z$ cross sections. Panels (a,b,c) present vorticity magnitude $\omega = |\nabla \times \mathbf{u}|$, (d,e,f) velocity divergence $\nabla \cdot \mathbf{u}$, and (g,h,i) normalized density-gradient magnitude $|\nabla \rho| / \rho$.}
    \label{fig:field_plot_sub}
\end{figure*}

\begin{figure*}
    \centering
    \includegraphics[width=0.95\textwidth]{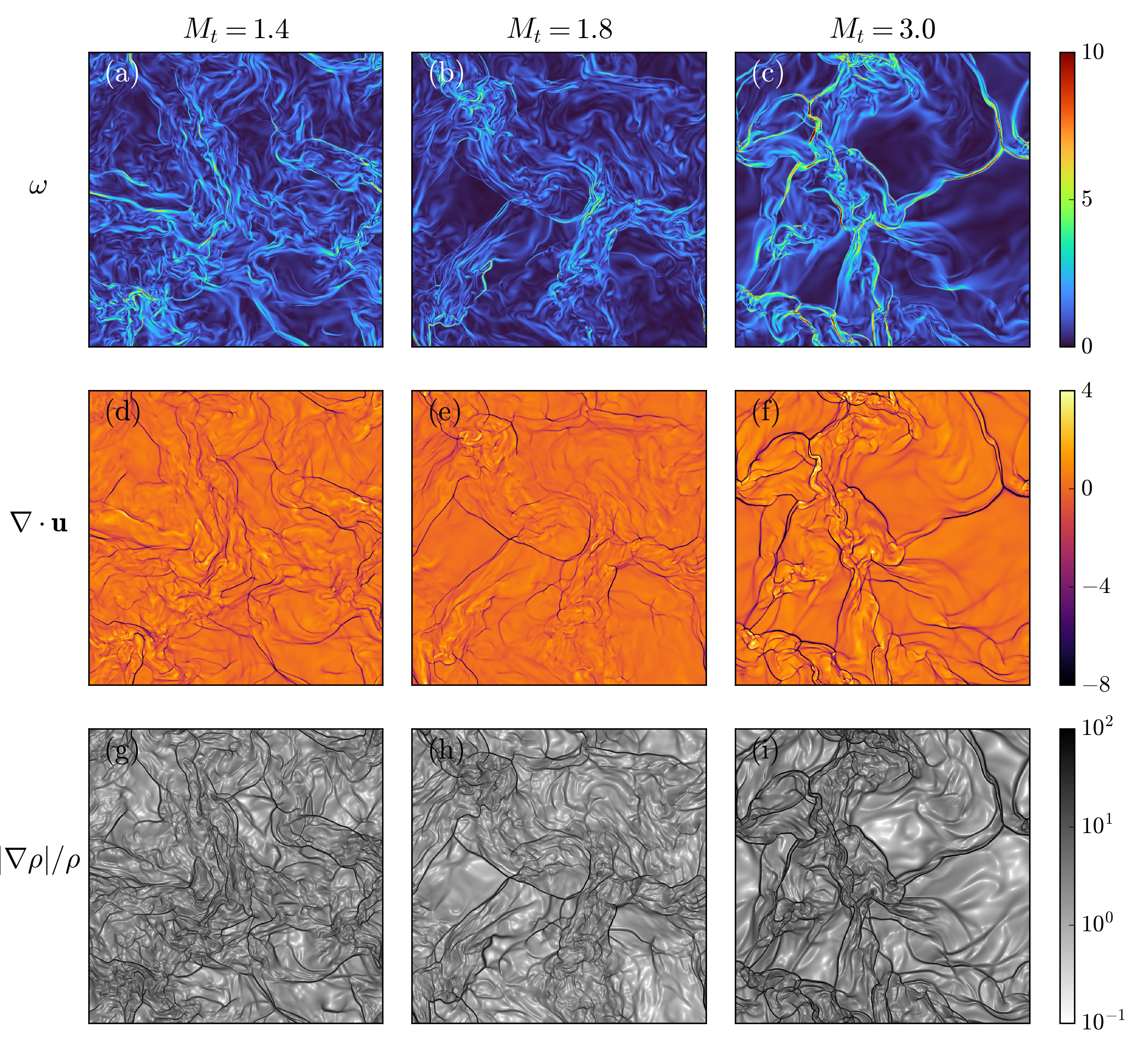}
    \caption{Flow structures in supersonic turbulence at $M_t = 1.4$ (left), $1.8$ (middle), and $3.0$ (right), shown on $x$–$z$ cross sections. Panels (a,b,c) present vorticity magnitude $\omega = |\nabla \times \mathbf{u}|$, (d,e,f) velocity divergence $\nabla \cdot \mathbf{u}$, and (g,h,i) normalized density-gradient magnitude $|\nabla \rho| / \rho$.}
    \label{fig:field_plots_super}
\end{figure*}

All simulations were integrated for more than 40 eddy turnover times. For our analysis, we discard the initial transients and focus on the statistically stationary state. Figure~\ref{fig:steady} presents the time series of the total kinetic energy, $E_u$, and the maximum local Mach number, $M_{t,\mathrm{max}}$, over 20 eddy turnover times within this stationary regime. As shown, $E_u$ remains nearly stable, whereas $M_{t,\mathrm{max}}$ exhibits large fluctuations above the mean. These deviations reflect intermittent bursts of intense compression and strong localized shocks.

Figures~\ref{fig:field_plot_sub} and \ref{fig:field_plots_super} display instantaneous $x$--$z$ cross-sections of the vorticity magnitude $\omega = |\nabla\times\mathbf{u}|$ (top row), velocity divergence $\nabla\cdot\mathbf{u}$ (middle row), and normalized density gradient magnitude $|\nabla\rho|/\rho$ (bottom row) in the statistically stationary state. Figure~\ref{fig:field_plot_sub} corresponds to the subsonic and transonic cases ($M_t = 0.2, 0.7, 1.0$), while Fig.~\ref{fig:field_plots_super} presents the supersonic runs ($M_t = 1.4, 1.8, 3.0$); for movies, see Ref.~\cite{movie}. The divergence fields highlight regions of intense compression, where large negative values of $\nabla \cdot \mathbf{u}$ correspond to local Mach numbers significantly exceeding $M_t$. These sharp discontinuities are similarly captured by the density gradients, which effectively resolve the widths of the simulated shocks.

The progression from $M_t=0.2$ to $M_t=3.0$ demonstrates the effects of turbulent Mach number and forcing composition on the structural evolution of compressible turbulence. As $M_t$ increases, the magnitudes of both $\nabla\cdot\mathbf{u}$ and $|\nabla\rho|/\rho$ rise significantly, signalling a progressive strengthening of the shock structures. In the nearly incompressible limit ($M_t = 0.2$), such structures are virtually absent. However, in the subsonic and transonic cases, the flow is characterized by fragmented shocklets distributed throughout the domain. In the supersonic simulations, these features evolve into strong, persistent shock sheets. The density gradients further highlight these sharp discontinuities—representing the numerical shock widths—where density jumps of several orders of magnitude occur within just a few grid cells. These structures are consistent with previous investigations of high-Mach-number turbulence~\citep{Kritsuk:ApJ2007,Federrath:MNRAS2013}.

In the subsonic regime, the vorticity field is filled by fine-scale structures distributed throughout the domain, a consequence of the high Reynolds numbers ($Re_\lambda$). As $M_t$ increases, however, the spatial organization changes. Notably, regions of intense vorticity become increasingly concentrated in the vicinity of shock structures.  This localization arises because vorticity is generated via baroclinic torque at curved shock fronts and subsequently amplified by vortex stretching, where solenoidal cascades develop from compressive seeds~\cite{Lele:ARFM1994, Federrath:MNRAS2013}.

\subsection{Energy spectra}

To compute energy spectra and fluxes, we employ the mode-to-mode formalism developed in~\cite{Singh:PRF2025}. Using the Helmholtz decomposition, we decompose the velocity field into rotational ($\alpha = R$) and compressible ($\alpha = C$) components, denoted by $\mathbf{u}_{\alpha}(\mathbf{k})$. The corresponding density-weighted velocity fields are defined as  $\mathbf{v}_{\alpha}(\mathbf{k})$, where $\mathbf{v}(\mathbf{x}) = \rho(\mathbf{x}) \, \mathbf{u}(\mathbf{x})$. The kinetic energy for each mode is given by
\begin{equation}
E_{\alpha}(\mathbf{k}) = \frac{1}{2} \mathrm{Re}\big[ \mathbf{v}_{\alpha}(\mathbf{k}) \cdot \mathbf{u}_{\alpha}^*(\mathbf{k}) \big],
\end{equation}
yielding the total kinetic energy spectrum
\begin{equation}
E_u(\mathbf{k}) = E_R(\mathbf{k}) + E_C(\mathbf{k}).
\end{equation}
In addition to the kinetic energy spectra, the density spectrum is defined as~\cite{Wang:PRF2017,Wang:PRE2018}
\be
E_\rho({\bf k}) = \frac{1}{2} |\rho({\bf k})|^2.
\ee
We compute the kinetic energy (KE) spectra for the rotational and compressive velocity components, $E_R(k)$ and $E_C(k)$, the total kinetic energy spectrum $E_u(k)$, and the density spectrum $E_\rho(k)$, averaged over 20 snapshots. Figure~\ref{fig:spectra_flux_sub} presents these spectra for $M_t = 0.2$, $0.7$ and $1.0$, while Fig.~\ref{fig:spectra_flux_super} displays the results for $M_t = 1.4, 1.8$, and $3.0$. The errors in the spectral exponents are less than $1\%$. In the nearly incompressible limit ($M_t = 0.2$), both kinetic energy and density spectra scale as $k^{-5/3}$, consistent with the results of ~\citet{Zank:PF1991}. For the subsonic and transonic cases ($M_t = 0.7$ and $1.0$), the rotational spectrum retains Kolmogorov scaling ($E_R(k) \sim k^{-5/3}$), whereas the compressive component exhibits a Burgers-like $k^{-2}$ scaling. The density spectrum similarly follows this $k^{-2}$ behavior. These observations are consistent with established findings for subsonic turbulence~\cite{Kida:JSC1990,Wang:PRL2013,Schmidt:PRE2019,Sakurai:PF2024,Wang:PRE2018}.

In contrast, the energy distribution changes significantly in the supersonic regime ($M_t = 1.4, 1.8$, and $3.0$). With increasing $M_t$, $E_R(k)$ steepens, deviating from the Kolmogorov $k^{-5/3}$ law and approaching a Burgers-like $k^{-2}$ slope at $M_t = 3.0$, despite the use of predominantly solenoidal forcing. Furthermore, while $E_C(k)$ follows $k^{-2}$ at $M_t = 1.4$, it becomes progressively shallower at higher Mach numbers. These results contrast with the shallower scaling for both rotational and compressive KE ($E_\alpha(k) \sim k^{-3/2}$) reported by \citet{Kritsuk:ApJ2007} and \citet{Federrath:AA2010}. We note, however, that those studies defined the kinetic energy as $w^2/2$, where $\mathbf{w} = \sqrt{\rho}\,\mathbf{u}$ is the density-weighted velocity. The density spectrum also flattens with increasing $M_t$, approaching $k^{-1}$ at $M_t = 3.0$, which agrees with high Mach number results in the literature~\cite{Kritsuk:ApJ2007, Federrath:AA2010}. In all cases, the total kinetic energy remains close to the solenoidal spectrum. These results demonstrate the stark difference in energy distribution at different scales between supersonic and subsonic turbulence. 


\subsection{Energy transfers and fluxes}\label{subsec:ET}
Now, we report the various energy fluxes in the flow and assess the effect of compressibility in energy cascade. Following the formalism of \citet{Singh:PRF2025}, the spectral evolution of the kinetic energy for the rotational ($\alpha=R$) and compressive ($\alpha=C$) modes is governed by
\bea
\partial_t E_\alpha(\mathbf{k})  &=&   \sum_{\bf p} S^{\alpha \alpha}({\bf k|p|q}) + \sum_{\bf p} S^{\alpha \beta}({\bf k|p|q}) \nonumber \\     && - Q_{I, \alpha}(\mathbf{k})
 - D_{I, \alpha}(\mathbf{k})
 + \mathcal{F}_{\alpha}(\mathbf{k}),
\eea 
where ${\bf k =p+q}$. The first term on the right-hand side is the \textit{mode-to-mode energy transfer }from ${\bf u}_\alpha({\bf p})$ to ${\bf u}_\alpha({\bf k})$. It represents the non-linear exchange of energy within the rotational modes (via $S^{RR}$) or within the compressive modes (via $S^{CC}$). 
The interaction between different kinetic energy components is described by the cross-transfer term $S^{\alpha \beta}$, which quantifies the energy exchanged between rotational and compressive modes. Additionally, ${\bf u}_\alpha({\bf k})$ exchanges energy with internal energy via pressure work $Q_{I,\alpha}(\mathbf{k})$ and viscous dissipation $D_{I, \alpha}(\mathbf{k})$.  $\mathcal{F}_{\alpha}(\mathbf{k})$ is the kinetic energy injection rate by the external force component. These terms are defined as 
\begin{flalign}
    & S^{\alpha \alpha}(\mathbf{k}|\mathbf{p}|\mathbf{q}) = \frac{1}{2} \mathrm{Im}\big[\{\mathbf{k} \cdot \mathbf{u}(\mathbf{q})\} \{\mathbf{v}_{\alpha}(\mathbf{p}) \cdot \mathbf{u}^*_\alpha (\mathbf{k})\} + \nonumber && \\
    & \hspace{9em} \{\mathbf{p} \cdot \mathbf{u}(\mathbf{q})\}\{\mathbf{u}_{\alpha}(\mathbf{p}) \cdot \mathbf{v}^*_\alpha (\mathbf{k})\}\big], \label{eq:Suu_alpha_alpha_u} && \\
    & S^{\alpha \beta}(\mathbf{k}|\mathbf{p}|\mathbf{q}) = \frac{1}{2} \mathrm{Im}\big[\{\mathbf{k} \cdot \mathbf{u}(\mathbf{q})\} \{\mathbf{v}_{\beta}(\mathbf{p}) \cdot \mathbf{u}^*_\alpha (\mathbf{k})\} + \nonumber && \\
    & \hspace{9em} \{\mathbf{p} \cdot \mathbf{u}(\mathbf{q})\}\{\mathbf{u}_{\beta}(\mathbf{p}) \cdot \mathbf{v}^*_\alpha (\mathbf{k})\}\big], && \\
    & Q_{I,R}(\mathbf{k}) = \frac{1}{2} \mathrm{Re}\big[\tilde{\mathbf{p}}(\mathbf{k}) \cdot \mathbf{v}_R^*(\mathbf{k})\big], && \\
    & Q_{I,C}(\mathbf{k}) = \frac{1}{2} \mathrm{Re}\big[\tilde{\mathbf{p}}(\mathbf{k}) \cdot \mathbf{v}_C^*(\mathbf{k})\big] - \frac{1}{2} \mathrm{Im}\big[p(\mathbf{k}) \{\mathbf{k} \cdot \mathbf{u}_C^*(\mathbf{k}) \}\big], && \\
    & D_{I,\alpha}(\mathbf{k}) = \frac{1}{2} \mathrm{Re}\big[\mathbf{d}_{\alpha}(\mathbf{k}) \cdot \mathbf{u}_{\alpha}^*(\mathbf{k}) + \tilde{\mathbf{d}}_{\alpha}(\mathbf{k}) \cdot \mathbf{v}_{\alpha}^*(\mathbf{k}) \big], && \\
    & \mathcal{F}_{\alpha}(\mathbf{k}) = \frac{1}{2} \mathrm{Re}\big[\mathbf{F'}_{\alpha}(\mathbf{k}) \cdot \mathbf{u}_{\alpha}^*(\mathbf{k}) + \mathbf{F}_{\alpha}(\mathbf{k}) \cdot \mathbf{v}_{\alpha}^*(\mathbf{k}) \big]. &&
\end{flalign}

In these expressions, $p(\mathbf{k})$ represents the Fourier amplitude of the pressure, and $\mathbf{F}_{\alpha}(\mathbf{k})$ denotes the Fourier amplitudes of the rotational and compressible forcing components. The remaining quantities are defined as
\be
\tilde{\mathbf{p}} = \nabla p / \rho, ~ \mathbf{d} = -\partial_j \tau_{ij}, ~ \tilde{\mathbf{d}} = \mathbf{d} / \rho,~ \text{and} ~ \mathbf{F}' = \rho \mathbf{F},
\ee
where $\tau_{ij}$ denotes the viscous stress tensor.

We define different energy fluxes corresponding to these transfer terms. To quantify interscale energy transfer, we define the pure flux
\begin{equation}
\Pi_{\alpha}(K) = \sum_{k > K} \sum_{p \le K} S^{\alpha \alpha}(\mathbf{k|p|q}),
\end{equation}
which measures the net rate of energy transfer from $\alpha$-modes (rotational/compressible) within a wavenumber sphere of radius $K$ to $\alpha$-modes (rotational/compressible) outside the sphere. Physically, $\Pi_{R}$ represents the traditional Kolmogorov-like cascade where large-scale solenoidal eddies break down into smaller vortical structures. In contrast, $\Pi_{C}$ represents the acoustic or shock-like cascade, where large-scale compressions steepen into small-scale shocklets or shock sheets. Similarly, the cross-flux
\begin{equation}
\Pi^{\alpha <}_{\beta}(K) = \sum_{k > K} \sum_{p} S^{\alpha \beta}(\mathbf{k|p|q}),
\end{equation}
represents the net transfer from $\alpha$-modes inside the sphere to all $\beta$-modes, allowing us to track exchanges between solenoidal and compressible motions. For example, $\Pi^{R<}_{C}$ quantifies how the deformation of rotational eddies acts as a source for new compressive structures at all scales. The energy exchanges between $E_u$ and $I$ are captured through fluxes associated with $Q$ and $D$:
\begin{equation}
\Pi^{\alpha <}_{I,Q}(K) = \sum_{k \le K} Q_{I,\alpha}(\mathbf{k}),
\end{equation}
\begin{equation}
\Pi^{\alpha <}_{I,D}(K) = \sum_{k \le K} D_{I,\alpha}(\mathbf{k}).
\end{equation}
We also define the rotational and compressive average dissipation rates as 
\begin{equation}
\epsilon_\alpha = \sum_{k \in K} D_{I,\alpha}(\mathbf{k}).
\end{equation}
For the definition of the mode-to-mode transfer functions and further details, see~\cite{Singh:PRF2025}, which form the basis of the flux definitions used in this study.




The fluxes are normalized using  
\be
\epsilon_T = \epsilon + \mathcal{W},
\ee
where $\epsilon$ is the average dissipation rate and $\mathcal{W} = \langle -p\,\nabla \cdot \mathbf{u} \rangle$ denotes the average pressure--dilatation. The quantity $\epsilon_T$ therefore equals the total energy injection rate ($\epsilon^{\mathrm{inj}}$) at steady state, or equivalently, the net rate of kinetic--to--internal energy conversion~\cite{Lele:ARFM1994}. We denote the normalized fluxes and energy injection rates with tildes (e.g., $\tilde{\Pi}_R(k)$ vs. ${\Pi}_R(k)$) to distinguish them from their unnormalized forms.

\begin{figure*}
    \centering
    \includegraphics[width=1\textwidth]{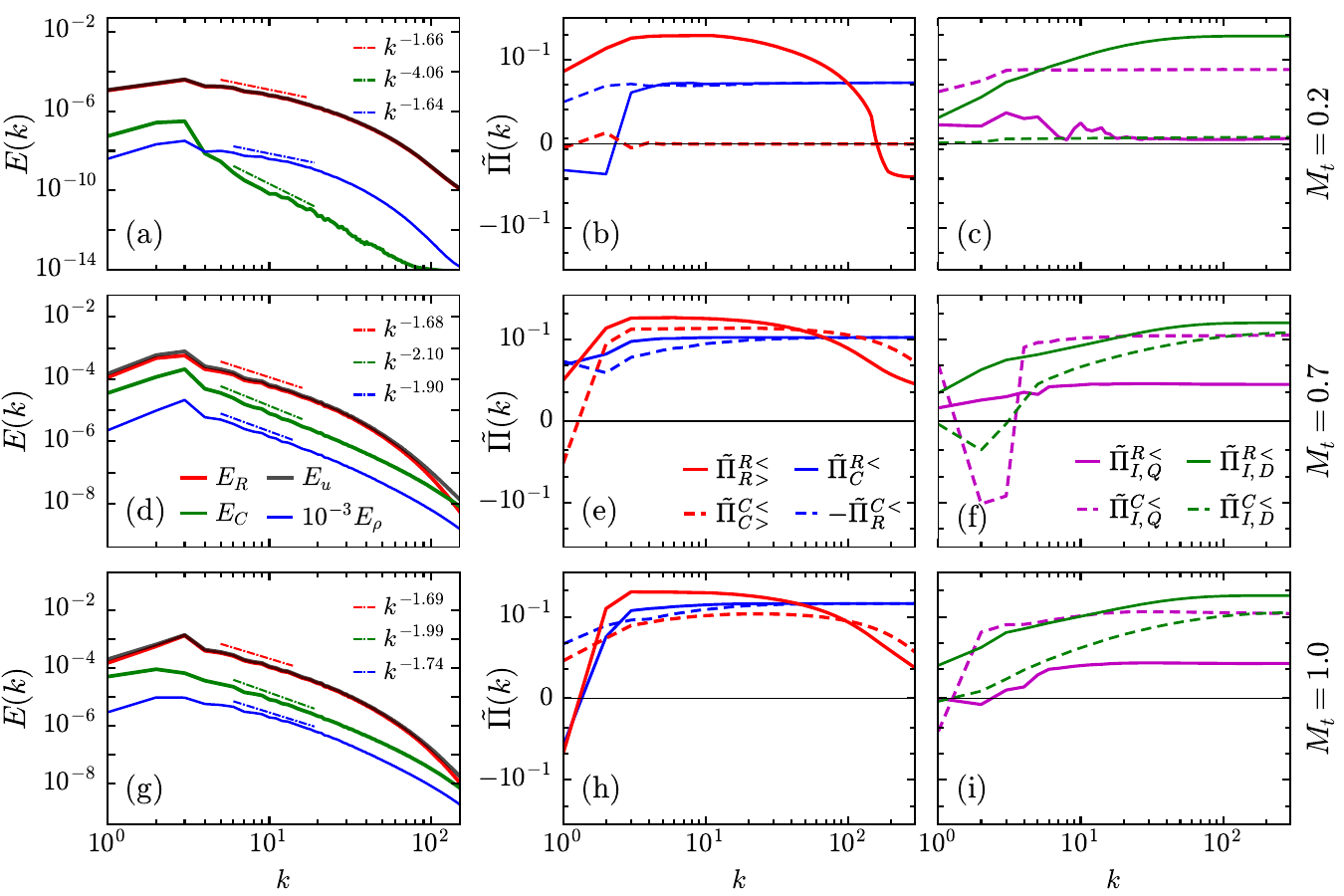}
    \caption{For $M_t = 0.2$ (top row), $M_t = 0.7$ (middle row), and (e,f) $M_t = 1.0$ (bottom row): (a,d,g) Plots of the time-averaged turbulent energy spectra for rotational kinetic energy $E_R$ (red), compressive kinetic energy $E_C$ (light green), total kinetic energy $E_u$ (gray), and density $E_{\rho}$ (blue). (b,e,h) Plots of time-averaged \textit{normalized} energy fluxes from the non-linear transfer term: $\tilde{\Pi}_{R}$ (solid red), $\tilde{\Pi}_{C}$ (dashed red), $\tilde{\Pi}^{R<}_{C}$ (solid blue), and $-\tilde{\Pi}^{C<}_{R}$ (dashed blue). (c,f,i) Plots of time-averaged \textit{normalized} energy fluxes due to pressure work and dissipation: $\tilde{\Pi}^{R<}_{I,Q}$ (solid magenta), $\tilde{\Pi}^{C<}_{I,Q}$ (dashed magenta), $\tilde{\Pi}^{R<}_{I,D}$ (solid green), and $\tilde{\Pi}^{C<}_{I,D}$ (dashed green).}
    \label{fig:spectra_flux_sub}
\end{figure*}

\begin{figure*}
    \centering
    \includegraphics[width=1\textwidth]{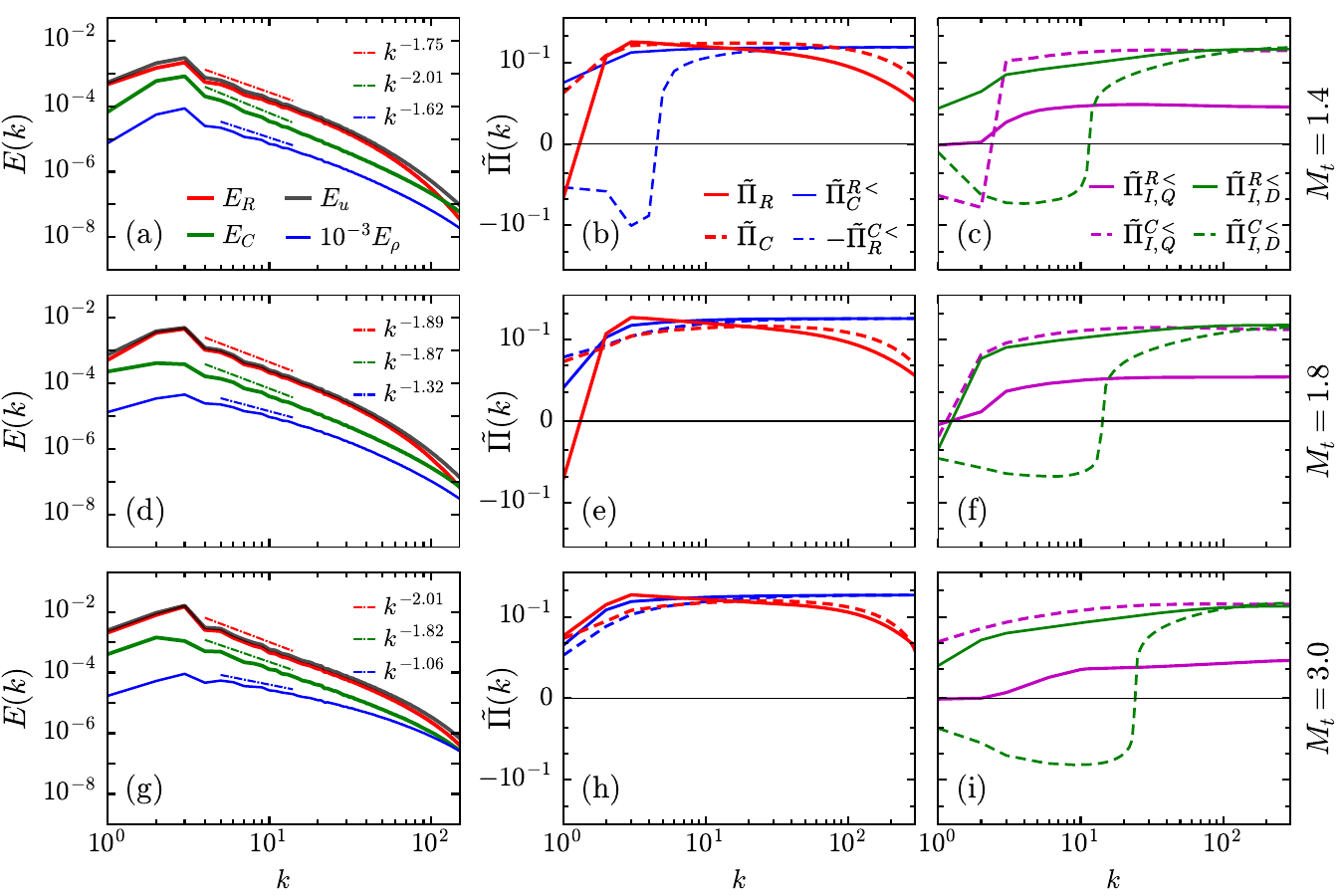}
    \caption{For $M_t = 1.4$ (top row), $M_t = 1.8$ (middle row), and (e,f) $M_t = 3.0$ (bottom row): (a,d,g) Plots of the time-averaged turbulent energy spectra for rotational kinetic energy $E_R$ (red), compressive kinetic energy $E_C$ (light green), total kinetic energy $E_u$ (gray), and density $E_{\rho}$ (blue). (b,e,h) Plots of time-averaged \textit{normalized} energy fluxes from the non-linear transfer term: $\tilde{\Pi}_{R}$ (solid red), $\tilde{\Pi}_{C}$ (dashed red), $\tilde{\Pi}^{R<}_{C}$ (solid blue), and $-\tilde{\Pi}^{C<}_{R}$ (dashed blue). (c,f,i) Plots of time-averaged \textit{normalized} energy fluxes due to pressure work and dissipation: $\tilde{\Pi}^{R<}_{I,Q}$ (solid magenta), $\tilde{\Pi}^{C<}_{I,Q}$ (dashed magenta), $\tilde{\Pi}^{R<}_{I,D}$ (solid green), and $\tilde{\Pi}^{C<}_{I,D}$ (dashed green).}
    \label{fig:spectra_flux_super}
\end{figure*}

We compute the normalized energy fluxes, averaged over 15--20 snapshots at steady state. Figure~\ref{fig:spectra_flux_sub} (b,c,e,f,h,i) presents these fluxes for the subsonic and transonic regimes ($M_t = 0.2, 0.7, 1.0$), while Fig.~\ref{fig:flux_trend}(a--c) further illustrates the energy transfer pathways. In the nearly incompressible limit ($M_t = 0.2$), the rotational flux $\tilde{\Pi}_R(k)$ remains constant across the inertial range. The compressive flux $\tilde{\Pi}_C(k)$, cross-fluxes, and pressure dilatation are negligible, consistent with ideal incompressible dynamics. In the subsonic regime ($M_t = 0.7$), both $\tilde{\Pi}_R(k)$ and $\tilde{\Pi}_C(k)$ exhibit constant plateaus in the inertial range ($4 \lessapprox k \lessapprox 15$)~\cite{Wang:PRL2013,Wang:JFM2018}. The cross-transfer is weak ($\tilde{\Pi}^{R <}_{C}(k) \approx 0.1$) and confined to large scales. Consequently, it has a negligible impact on the inertial-range fluxes~\cite{Verma:JPA2022}. Interestingly, in the transonic case ($M_t = 1.0$), although the cross-flux strengthens significantly ($\tilde{\Pi}^{R <}_{C}(k) \approx 0.3$), it remains restricted to small wavenumbers. As a result, both $\tilde{\Pi}_R(k)$ and $\tilde{\Pi}_C(k)$ maintain their constant, cascade-like behavior in the inertial range even in the transonic case. For both $M_t = 0.7$ and $1.0$, pressure dilatation $\tilde{\Pi}^{C<}_{I,Q}(k)$ becomes comparable to the compressive viscous dissipation. However, it remains nearly constant at intermediate and large wavenumbers, indicating that energy transfer via pressure work occurs primarily at large scales~\cite{Aluie:PRL2011}. In contrast, the pressure contribution to the rotational component, $\tilde{\Pi}^{R<}_{I,Q}(k)$, remains negligible.

The energy transfer in supersonic flows differs significantly from that in the subsonic and transonic regimes. Figure~\ref{fig:spectra_flux_super}(b,c,e,f,h,i) presents these fluxes for supersonic turbulence, with rows corresponding to $M_t = 1.4$, $1.8$, and $3.0$. Additionally, Fig.~\ref{fig:flux_trend}(d--f) illustrates the energy transfer pathways. The key features of the normalized fluxes are as follows:
\begin{itemize}
    \item \textit{Rotational component}: Energy injected into rotational modes, $\tilde{\epsilon}^{\mathrm{inj}}_R$, splits into two pathways: a forward cascade, $\tilde{\Pi}_{R}$, and a cross-transfer to compressible modes, $\tilde{\Pi}^{R<}_{C}$. In contrast to the subsonic case, a significant fraction of $\tilde{\epsilon}^{\mathrm{inj}}_R$ is transferred to the compressible modes in supersonic turbulence. Although this transfer is most active at large scales, $\tilde{\Pi}^{R<}_{C}$ continues to grow across the intermediate scales, indicating substantial cross-transfer within the inertial range. Consequently, the energy available for the forward cascade diminishes, causing $\tilde{\Pi}_R$ to decrease with $k$. This behavior contrasts sharply with the Kolmogorov-like constant flux observed in subsonic turbulence. As a result, the rotational spectrum $E_R(k)$ steepens, deviating from $k^{-5/3}$ scaling. Figure~\ref{fig:flux_trend} confirms that increasing $M_t$ amplifies $\tilde{\Pi}^{R<}_{C}$, driving the steepening of both $\tilde{\Pi}_{R}$ and $E_R(k)$. We also note that $\tilde{\Pi}^{R<}_{C}$ depends on the forcing parameter $\zeta$, remaining lower for mixed forcing than for predominantly solenoidal forcing at similar Mach numbers.\\
    
    \item \textit{Compressive component}: Figure~\ref{fig:flux_trend} shows that the compressible modes gain energy from two sources: direct forcing, $\tilde{\epsilon}^{\mathrm{inj}}_C$, and cross-transfer from rotational modes, $-\tilde{\Pi}^{C<}_{R}$. This energy splits into two paths: (a) a forward cascade to smaller scales via $\tilde{\Pi}_{C}(k)$, and (b) conversion to internal energy via pressure dilatation, $\tilde{\Pi}^C_{I,Q}(k)$. As seen in Fig.~\ref{fig:spectra_flux_super}(b,e,h), $\tilde{\Pi}_{C}(k)$ increases slightly across the inertial range. This occurs because the cross-transfer adds energy even at intermediate scales, increasing the energy available for the forward transfer. Consequently, $\tilde{\Pi}_{C}(k)$ rises across the inertial range. This leads to a shallower compressive spectrum, particularly at higher $M_t$. However, the rise in $\tilde{\Pi}_{C}(k)$ is much weaker than the steepening of $\tilde{\Pi}_{R}(k)$ because a portion of the transferred energy is converted to internal energy via pressure dilatation.\\

\begin{figure*}
\centering
\includegraphics[width=0.94\textwidth]{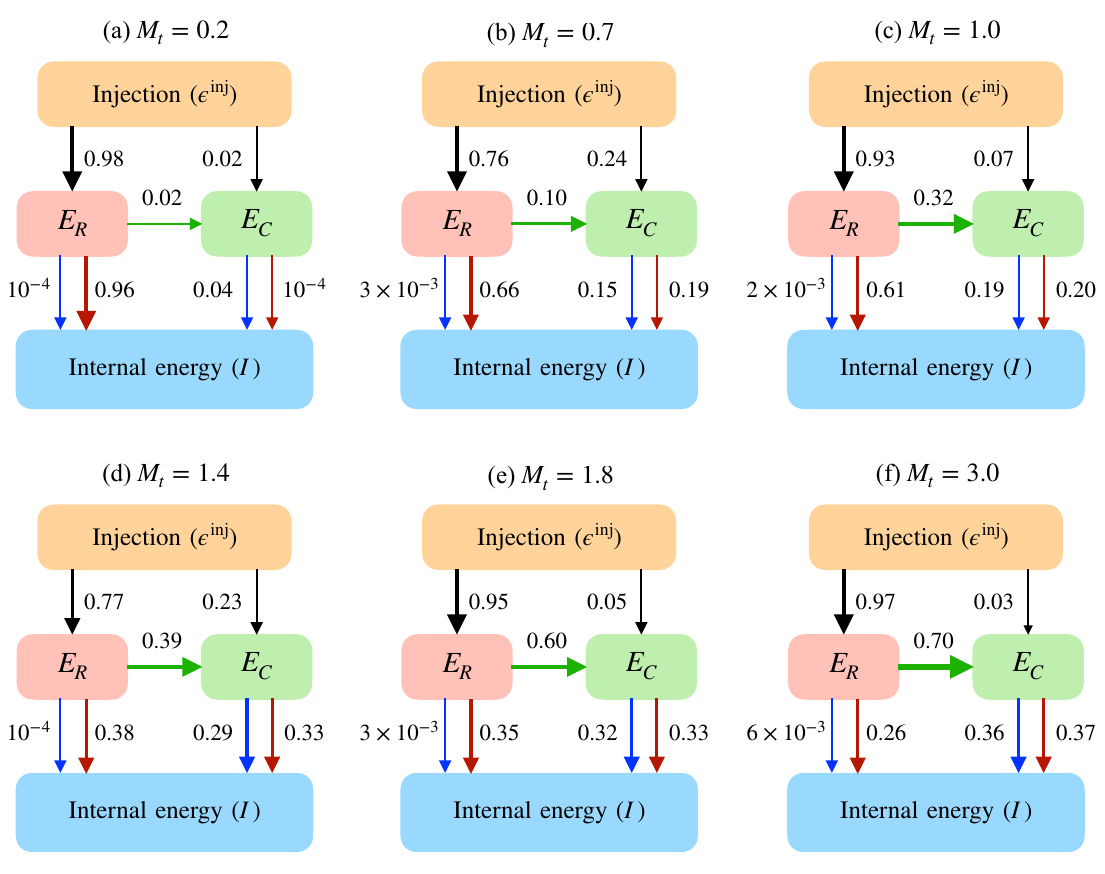}
\caption{For forced compressible turbulence with (a) $M_t = 0.2$, (b) $M_t = 0.7$, (c) $M_t = 1.0$, (d) $M_t = 1.4$, (e) $M_t = 1.8$, (f) $M_t = 3.0$: Schematic diagrams exhibiting various energy transfers. The normalized energy injection rates $\tilde{\epsilon}^\mathrm{inj}_R, \tilde{\epsilon}^\mathrm{inj}_C$ (black arrows); maxima of $\tilde{\Pi}^{R<}_{C}(k)$ (green arrows), maxima of $\tilde{\Pi}^{C<}_{I,Q}(k), \tilde{\Pi}^{R<}_{I,Q}(k)$ (blue arrows); total normalized dissipation rates $\tilde{\epsilon}_R, \tilde{\epsilon}_C$ (red arrows).}
\label{fig:flux_trend}
\end{figure*}

    \item \textit{Pressure dilatation}: Pressure dilatation converts kinetic energy to internal energy via pressure work. Figure~\ref{fig:flux_trend} shows that this work is significant for the compressive component and remains comparable to the compressive viscous dissipation. Notably, the compressive pressure dilatation increases with increasing $M_t$. Figure~\ref{fig:spectra_flux_super}(c,f,i) reveals that the compressive pressure dilatation flux, $\tilde{\Pi}^{C <}_{I,Q}(k)$, is most active at large scales (small $k$). However, unlike the subsonic case, it retains a non-negligible magnitude throughout the inertial range. In contrast, the pressure flux for the rotational component vanishes ($\tilde{\Pi}^{R <}_{I,Q} \rightarrow 0$). This aligns with incompressible theory, where pressure does not contribute to energy transfers~\cite{Orszag:CP1973,Verma:PR2004,Verma:book:ET}.\\

    \item \textit{Viscous dissipation}: Viscous dissipation irreversibly converts kinetic energy into internal energy. The maximum of
    viscous flux, $\tilde{\Pi}^{\alpha <}_{I,D}(k)$, matches the total dissipation rate $\tilde{\epsilon}_\alpha$. As shown in Fig.~\ref{fig:spectra_flux_super}(c,f,i), $\tilde{\Pi}^{\alpha <}_{I,D}(k)$ increases gradually with wavenumber $k$ for each component. This indicates that dissipation is active across all scales, though predominantly at small scales. Interestingly, the compressive dissipation spectrum exhibits small negative values at low wavenumbers, implying a reverse energy transfer from internal to compressive modes. This behavior arises from the density-weighted formulation of the viscous divergence~\cite{Singh:PRF2025} and intensifies at higher $M_t$. Despite this local reversal, the total compressive dissipation remains positive. Figure~\ref{fig:flux_trend} demonstrates that as $M_t$ increases, the total rotational viscous dissipation decreases, whereas the compressive dissipation increases.
\end{itemize}

\subsection{Trend with forcing and turbulent Mach number}

\begin{table*}
\centering
\renewcommand{\arraystretch}{1.5} 
\setlength{\tabcolsep}{6pt}       
\caption{Numerically-computed normalized injection rates ($\tilde{\epsilon}^{\mathrm{inj}}_R$,$\tilde{\epsilon}^{\mathrm{inj}}_C$), energy ratios ($E_R/E_C$), normalized dissipation rates ($\tilde{\epsilon}_R$,$\tilde{\epsilon}_C$), maximum of normalized fluxes, and the scaling exponents $\alpha$ for various energy spectra, along with their trends with increasing turbulent Mach numbers ($M_t$). }
\vspace{4pt}
\begin{tabular}{lccccccc}
\hline
$M_t$ & 0.2 & 0.7 & 1.0 & 1.4 & 1.8 & 3.0 & Trend\\
\hline
$\tilde{\epsilon}^{\mathrm{inj}}_R$ & 0.98 & 0.76 & 0.93 & 0.77 & 0.95 & 0.97 & - \\
$\tilde{\epsilon}^{\mathrm{inj}}_C$ & 0.02 & 0.24 & 0.07 & 0.23 & 0.05 & 0.03 & - \\
$E_R/E_C$  & 322.3 & 4.28 & 10.81 & 2.98 & 6.96 & 6.12 & -\\
$\tilde{\epsilon}_R$ & 0.96 & 0.66 & 0.61 & 0.38 & 0.35 & 0.26 & Decreases\\
$\tilde{\epsilon}_C$ & $10^{-4}$ & 0.19 & 0.20 & 0.33 & 0.33 & 0.37 & Increases\\
$\tilde{\Pi}^{R}_{C}$ & 0.02 & 0.10 & 0.32 & 0.39 & 0.60 & 0.70 & Increases\\
$-\tilde{\Pi}^{C}_{R}$ & 0.02 & 0.10 & 0.32 & 0.39 & 0.60 & 0.70 & Increases\\
$\tilde{\Pi}^{R}_{I,Q}$ & $10^{-4}$ & $3 \times 10^{-3}$ & $2 \times 10^{-3}$ & $10^{-4}$ & $3 \times 10^{-3}$ & $6 \times 10^{-3}$ & $\approx 0$\\
$\tilde{\Pi}^{C}_{I,Q}$ & 0.04 & 0.15 & 0.19 & 0.29 & 0.32 & 0.36 & Increases\\

$\alpha_u$ & $-$1.66 & $-$1.73 & $-$1.70 & $-$1.80 & $-$1.87 & $-$1.96 & Steepens\\
$\alpha_R$ & $-$1.66 & $-$1.68 & $-$1.69 & $-$1.75 & $-$1.89 & $-$2.01 & Steepens\\
$\alpha_C$ & $-$4.06 & $-$2.10 & $-$1.99 & $-$2.01 & $-$1.87 & $-$1.82 & Shallows\\
$\alpha_\rho$ & $-$1.64 & $-$1.90 & $-$1.74 & $-$1.62 & $-$1.32 & $-$1.06 & Shallows\\
\hline
\end{tabular}
\label{table:flux}
\end{table*}

\begin{figure*}[htpb]
\centering \includegraphics[width=0.95\textwidth]{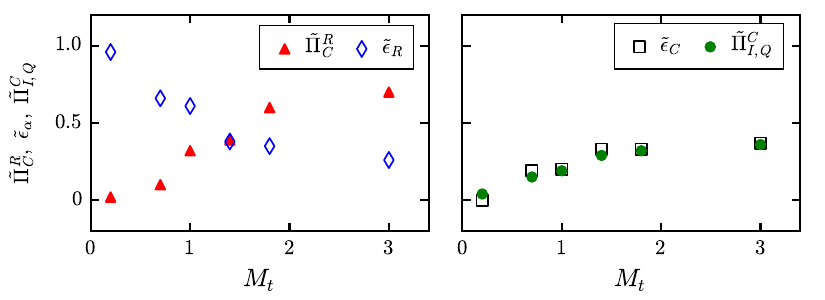}
\caption{Plots of normalized energy transfer terms with turbulent Mach number $M_t$: (a) rotational energy dissipation $\tilde{\epsilon}_R$ (blue diamonds) and maximum rotational-to-compressive cross-transfer $\tilde{\Pi}^R_C$ (red triangles); (b) compressive energy dissipation $\tilde{\epsilon}_C$ (hollow squares) and total pressure dilatation $\tilde{\Pi}^{C}_{I,Q}$ (green circles).}
 \label{fig:Mt_trend}
\end{figure*}

We summarize the variation of key turbulent statistics with turbulent Mach number ($M_t$) in Table~\ref{table:flux} and Fig.~\ref{fig:Mt_trend}. The simulations comprise two subsets: a primarily rotational forcing ($\zeta = 2/3$ for $M_t = 0.2, 1.0, 1.8, 3.0$) and a mixed forcing ($\zeta = 1/3$ for $M_t = 0.7, 1.4$). The normalized injection rates ($\tilde{\epsilon}^{\mathrm{inj}}_R, \tilde{\epsilon}^{\mathrm{inj}}_C$) show little variation with $M_t$ but depend strongly on the forcing parameter $\zeta$, which controls the ratio of rotational to compressive forcing amplitudes. For runs with $\zeta = 2/3$, the rotational injection rate $\tilde{\epsilon}^{\mathrm{inj}}_R$ ranges from $0.93$ to $0.98$, while for $\zeta = 1/3$, it drops to approximately $0.76$. Despite the relatively constant injection rates for a given $\zeta$, the flow composition changes significantly with increasing $M_t$. For instance, within the fixed $\zeta = 2/3$ subset, the ratio of rotational to compressive kinetic energy ($E_R/E_C$) drops from 322.3 to 6.12 as $M_t$ increases, indicating that compressibility becomes increasingly dominant as the Mach number rises.

While the energy injection at large scales is forcing-dependent, the subsequent energy transfers and spectral scaling exhibit a clear, monotonic dependence on $M_t$, indicating that these dynamics are predominantly governed by the Mach number rather than the forcing composition. Figure~\ref{fig:Mt_trend}(a) elucidates this trend. As $M_t$ increases, the normalized rotational dissipation $\tilde{\epsilon}_R$ decreases, while the cross-transfer flux from rotational to compressive modes, $\tilde{\Pi}^R_C$, rises markedly from 0.02 to 0.70. This confirms that in supersonic flows, a significant fraction of solenoidal energy is transferred into the compressive modes rather than being dissipated by viscosity. It is important to note from Fig.~\ref{fig:spectra_flux_sub} and Fig.~\ref{fig:spectra_flux_super} that while the fluxes $\Pi^{R<}_C$ and $-\Pi^{C<}_R$ do not strictly match at every scale, their total integrated transfers are identical (see Table~\ref{table:flux}), representing the net total transfer from rotational to compressive modes. For the compressive modes [Fig.~\ref{fig:Mt_trend}(b)], both the compressive dissipation $\tilde{\epsilon}_C$ and total compressive pressure dilatation $\tilde{\Pi}^C_{I,Q}$ increase with $M_t$, converting compressive kinetic energy into internal energy. Notably, $\tilde{\epsilon}_C$ and $\tilde{\Pi}^C_{I,Q}$ remain comparable across the entire Mach number range, with the exception of the near-incompressible limit ($M_t = 0.2$). These dynamical changes directly influence the spectral scaling exponents ($\alpha$) listed in Table~\ref{table:flux}, where the rotational and total velocity spectra steepen due to the energy drain via cross-transfer flux. In contrast, the compressive velocity spectrum shallows at higher Mach numbers. Consistent with previous studies~\cite{Kritsuk:ApJ2007,Federrath:AA2010}, the density spectrum approaches $k^{-1}$ at high $M_t$. It is worth noting that while these energy-cascade trends appear robust across our current setup, the specific energy-transfer rates may exhibit a stronger dependence on forcing in different parameter regimes.

\begin{figure*}
\centering \includegraphics[width=0.95\textwidth]{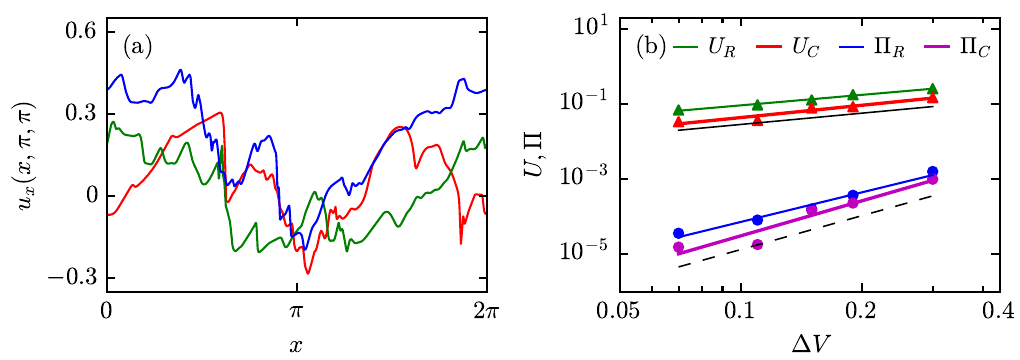}
\caption{(a) Plot of $u_x(x,y=\pi,z=\pi,t)$ vs.~$x$ at three different time snapshots at steady state for $M_t=3.0$ run. (b) Plots of root-mean-square velocities $U_R$ (green triangles) and $U_C$ (red triangles), inertial-range energy fluxes $\Pi_{R}$ (blue circles) and $\Pi_{C}$ (magenta circles) with $\Delta V$.  The solid black and dashed black lines represent  $\Delta V/\sqrt{12}$ and $\Delta V^3/(12L)$, respectively.}
 \label{fig:Burgers}
\end{figure*}

\subsection{Connection with Burgers turbulence}
Burgers equation is~\cite{Burgers:AAM1948} 
\be
\partial_t u + u \partial_x u = \nu \partial^2_{xx} u,
\ee
where $\nu$ is the kinematic viscosity. It
represents fully compressible fluid with no pressure term. In Burgers turbulence, the energy spectrum follows
\begin{equation}
E_B(k) = \frac{\langle (\Delta V)^2 \rangle}{2\pi L}k^{-2},
\label{eq:burger_eqn}
\end{equation}
where $\Delta V$ is the velocity jump across the shock and $L$ is the domain size~\cite{Saffman:book1968, Alam:PRF2022}.  \citet{Saffman:book1968} further established that the RMS velocity ($U_B$) and the energy flux ($\Pi_B$) scale with $\Delta V$ as
\bea
U_B &=& \frac{\Delta V}{\sqrt{12}},
\label{eq:UC_Burgers} \\
\Pi_B &=& \frac{(\Delta V)^3}{12L}.
\label{eq:PiC_Burgers} 
\eea
Now, we compare the above predictions with our numerical results.

In our simulation, we calculated $\Delta V$ by identifying velocity jumps across shocks in the $u_x$ component for a fixed $y$ and then averaging their magnitudes over a sample of 20 distinct spatial coordinates ($z$) and 20 time snapshots. This process provides good spatial and temporal averaging. This yield $\Delta V =  0.08$, 0.11, 0.15, 0.19 and 0.30 for $M_t = 0.7$, 1.0, 1.4, 1.8 and 3.0 respectively. The uncertainties in calculating $\Delta V$ is around 0.02. The estimates for $\Delta V$ are close to rms velocity ($U = cM_t$). For illustration, in Fig.~\ref{fig:Burgers}(a) we exhibit $u_x(x,y=\pi,z=\pi)$ for three different snapshots of the $M_t=3.0$ run. The figure shows that the velocity differences across shocks are of the order of $U$.

Figure~\ref{fig:Burgers}(b) presents the rotational and compressive RMS velocities ($U_R$, $U_C$ ) and the maximum energy fluxes ($\Pi_R$, $\Pi_C$). The compressive components align remarkably well with Burgers equation predictions (solid and dashed black lines). A best-fit analysis yields
\bea
U_C & =&  0.54 (\Delta V)^{1.08} \approx \frac{\Delta V}{\sqrt{12}}, \\
\Pi_C & = & 0.038\,(\Delta V)^{3.10} \approx \frac{(\Delta V)^3}{12L}.\label{eq:Pi_C}
\eea
Interestingly, the rotational component exhibits similar scaling trends with respect to the shock strength $\Delta V$:
\bea
U_R & =  & 0.79 (\Delta V)^{0.93}, \\
\Pi_R & = &  0.031 (\Delta V)^{2.62}.\label{eq:Pi_R}
\eea 

The physical mechanisms underlying these rotational scaling laws remain to be fully understood and need further investigation. The above discussion brings out interesting connections between the Burgers turbulence and compressible turbulence. We believe that Eqs.~(\ref{eq:UC_Burgers}, \ref{eq:PiC_Burgers}) can play a very important role in modelling complex astrophysical flows with high Mach numbers.

\section{Conclusions}\label{sec:conclusions}
While supersonic turbulence is of fundamental importance in astrophysical and engineering applications, a comprehensive understanding of its energy transfer remains incomplete. Previous numerical investigations have predominantly relied on Large Eddy Simulations (LES) or inviscid Euler simulations, often overlooking the full viscous dynamics. To address this, we performed Direct Numerical Simulations (DNS) of forced compressible turbulence across a wide range of turbulent Mach numbers, $M_t \in \{0.2, 0.7, 1.0, 1.4, 1.8, 3.0\}$ using a GPU-accelerated Python solver \texttt{DHARA}. By utilizing a seventh-order, low-dissipation TENO scheme, we were able to simultaneously capture fine-scale turbulent eddies and sharp shock fronts. Applying the formalism developed in \cite{Singh:PRF2025}, we calculated the energy spectra and fluxes for both subsonic and supersonic regimes. This work is a major advancement in understanding the mechanisms of energy transfers in compressible turbulence, particularly in the supersonic regime.

We conducted the simulations at a grid resolution of $1024^3$, varying the turbulent Mach number and ratio of rotational to compressive forcing. These high-resolution simulations reveal the distinct effect of the Mach number on flow structures and spectral scalings. We observe extremely thin, intense filaments identified as shock fronts, which occupy an increasingly larger volume fraction of the domain as the Mach number rises. The rotational velocity spectra steepen from a Kolmogorov-like $k^{-5/3}$ in the subsonic regime to a Burgers-like $k^{-2}$ in supersonic turbulence. Conversely, the compressive velocity spectrum becomes shallower at higher Mach numbers, deviating from standard Burgers scaling. For all cases, the total kinetic energy spectrum follows the scaling of the rotational component. These findings contrast with the $k^{-3/2}$ scaling reported in earlier supersonic studies (e.g., \citet{Kritsuk:ApJ2007} and \citet{Federrath:AA2010}), which relied on the density-weighted velocity $\mathbf{w} = \sqrt{\rho} \, \mathbf{u}$.

The transition from subsonic to supersonic flow regimes also reveals a fundamental change in the nature of inter-scale energy transfer. In subsonic flows, both rotational and compressive energy fluxes remain constant across the inertial range, with only a weak cross-transfer of energy from solenoidal to compressive modes. However, as the turbulent Mach number $M_t$ increases, this cross-transfer becomes dominant, extending well into the inertial range. This significant energy transfer results in a decline of the rotational flux across the inertial scales. This leads to the steepening of the rotational energy spectra observed at higher $M_t$. Conversely, the compressive flux increases within the inertial range as it receives energy via this enhanced cross-transfer, leading to the shallowing of the compressive spectra. Furthermore, pressure dilatation, which remains confined to large scales in subsonic turbulence, becomes non-negligible throughout the inertial range in supersonic flows. Interestingly, even though the injection rate at large scales depends on the forcing composition, the overall pattern of the energy cascade changes little with it, demonstrating that the inertial-range dynamics are predominantly governed by $M_t$.

Finally, we identify distinct scaling laws for the compressive component: the root-mean-square compressive velocity scales as $U_C \approx \Delta V/\sqrt{12}$, while the compressive energy flux follows $\Pi_C \approx (\Delta V)^3/(12 L)$, where $\Delta V \approx U$ (rms velocity). These scaling relations demonstrate a striking similarity between the compressive modes in supersonic turbulence and classical Burgers turbulence. These findings provide a robust theoretical basis for modeling the energy dynamics of extreme astrophysical flows, where compressibility plays a governing role.

Our findings significantly advance the understanding of energy transfer in supersonic flows. We demonstrate that supersonic turbulence cannot be described by a simple Kolmogorov- or Burgers-like cascade. Instead, the interplay of shocks, intermodal energy conversion, and pressure–velocity coupling fundamentally reshapes the energy cascade, significantly altering both the spectra and fluxes. These insights can be leveraged to develop reduced models to predict the observed spectral and flux behaviors—an essential step toward a predictive theory of supersonic turbulence in both astrophysical and engineering contexts. We aim to pursue this direction in future work.

\begin{acknowledgments}
The authors thank Sanjiva Lele, Katepalli Sreenivasan, Hussein Aluie, Hang Song, Lekha Sharma, Siddharth Rana, Shashwat Nirgudkar, Abhay Kumar, Abhishek Jha, and Manthan Verma for useful discussions. The authors thank Argonne Leadership Computing Facility (ALCF) and Oak Ridge Leadership Computing Facility (OLCF) for computer time through the Director's discretionary program. Simulations were performed on Polaris, Sophia, Frontier, HPC cluster of Kotak School of Sustainability (KSS), IIT Kanpur and our laboratory GPUs. Part of this work was done in the Centre for Turbulence Research, Stanford University, where MKV was a Visiting Senior Fellow. Part of this work was supported by KSS, IIT Kanpur grant (DORA /DORA/2023508I), Anusandhan National Research Foundation, India (SERB/PHY/2021522 and SERB/PHY/2021473), and the J. C. Bose Fellowship (SERB/PHY/2023488). 
\end{acknowledgments}

\appendix
\section{DNS code \texttt{\texttt{DHARA}}}\label{sec:dhara}

The numerical methods described in the \S\ref{sec:method} are implemented within the high-performance fluid solver \texttt{\texttt{DHARA}}~\cite{Tiwari:PNAS2025}, designed for both CPU and GPU architectures, which leverages Python for its flexibility and extensive scientific computing ecosystem. The solver is built with a strong emphasis on vectorized operations and memory efficiency to handle large-scale simulations effectively. The core numerical operations are performed using NumPy~\cite{numpy} for CPU-based computations, providing efficient array manipulation and mathematical functions. For accelerated computations on GPUs, the solver seamlessly integrates CuPy~\cite{cupy_learningsys2017}, a NumPy-compatible array library that utilizes NVIDIA CUDA. This dual-backend approach allows the same high-level Python code to execute efficiently on different hardware, providing significant performance gains on GPU-enabled systems. To further optimize performance on GPUs, custom CuPy \texttt{ElementwiseKernel} is employed for computationally intensive operations, ensuring maximum throughput. For parallelization, we use the mpi4py~\cite{mpi4py} library for both CPU and GPU.

\begin{figure*} 
    \centering
    \includegraphics[width = 0.75\textwidth]{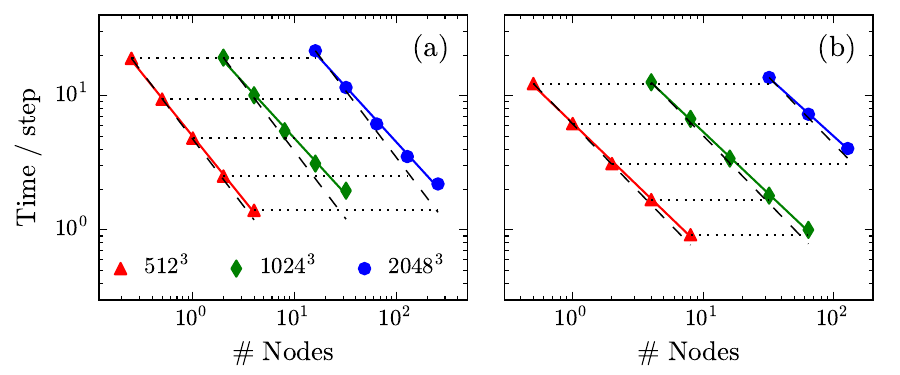}
    \caption{Scalability of \texttt{DHARA} on GPUs of (a) Frontier and (b) Polaris for 3D decaying compressible turbulence using TENO7 reconstruction. The figure shows the time taken per timestep vs. the number of nodes, demonstrating strong and weak scaling.}
    \label{fig:scaling}
\end{figure*}

To assess the performance of \texttt{\texttt{DHARA}}, we performed scaling analysis of code on AMD MI250X GPUs of Frontier (OLCF) and NVIDIA A100 GPUs of Polaris (ALCF). Each node of Frontier contains four AMD MI250X GPUs, each with two Graphics Compute Dies (GCDs), and each Polaris node contains four NVIDIA A100 GPUs. On both Frontier and Polaris, we conducted scaling tests for three-dimensional decaying compressible turbulence using the TENO7 reconstruction scheme. The grid size was varied from $512^3$ to $2048^3$, with the number of nodes increased from 1 to 256 on Frontier and from 1 to 128 on Polaris. Figure~\ref{fig:scaling}(a, b) shows the average time per timestep as a function of the number of nodes on Frontier and Polaris, respectively. The reported time is averaged over several timesteps. We observe that the time taken $T \propto n^{-1}$, where $n$ is the number of nodes, thus indicating strong scaling for \texttt{DHARA} in both systems. In addition, the code shows good weak scaling because the time taken remains unchanged when the grid size and number of nodes are increased proportionally. This demonstrates that \texttt{DHARA} achieves high scalability and efficient utilization of GPU resources across diverse architectures.

We performed several benchmark studies, including the well-known Taylor-Green vortex (TGV), isentropic vortex, and Kelvin-Helmholtz instability to assess the accuracy of the different reconstruction schemes implemented in our compressible fluid solver, \texttt{\texttt{DHARA}}. Briefly, the reconstruction schemes evaluated in this study include:
\begin{itemize}
    \item \textit{Linear}~\cite{Kurganov:JCP2000}: Using linear-reconstruction from left and right slopes.
    \item \textit{CWENO3-Z}~\cite{Levy:SJSC2000,Kurganov:JSC2000}: Third-order variant of Central WENO-Z, using a nonlinear blend of one quadratic polynomial and two linear polynomials.
    \item \textit{WENO5-Z, WENO7-Z}~\cite{Borges:JCP2008}: Fifth- and seventh-order variants of WENO-Z.
    \item \textit{TENO5, TENO7}~\cite{Fu:JCP2016}: Fifth- and seventh-order variants of TENO.
\end{itemize}

\subsection{Supersonic Taylor-Green vortex}\label{sec:tgv}

\begin{figure*}
    \centering
    \includegraphics[width = 0.85\textwidth]{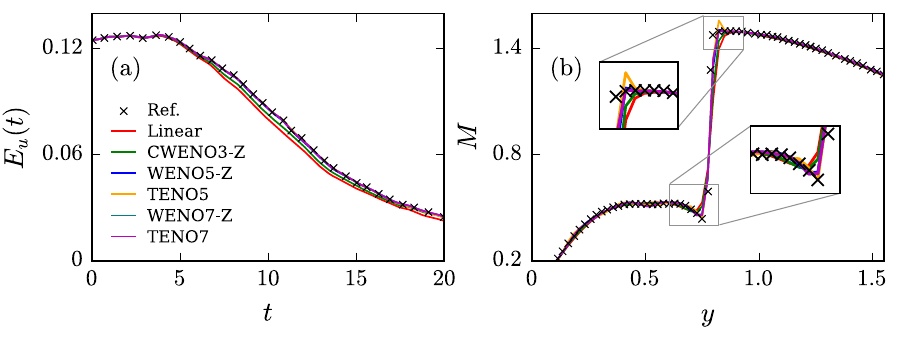}
    \caption{Case study using supersonic Taylor–Green vortex simulation. (a) Time evolution of the volume-averaged kinetic energy $E_u(t)$ for different reconstruction schemes. (b) Instantaneous Mach number profile $M$ along the $y$-axis at $x = z = 0$ at time $t = 2.5$. The reference points (Ref.) in both subplots correspond to a benchmark solution computed at $2048^3$ resolution from~\citet{Chapelier:PF2024}.}
    \label{fig:tgv}
\end{figure*}

The 3D Taylor-Green vortex (TGV) problem~\cite{Brachet:JFM1983} is considered in the supersonic regime~\cite{Lusher:AIAA2021,Chapelier:PF2024} without external forcing. The domain is $[-\pi, \pi]^3$ with periodic boundary conditions in all directions. We consider $M_0 = 1.25$, $\gamma = 1.4$, $\mathrm{Re}_0=1600$, $\mathrm{Pr}=0.71$ and grid size of $256^3$ for all runs. We track the time evolution of the volume-averaged kinetic energy $E_u$, which is plotted in Fig.~\ref{fig:tgv} (a) for various reconstruction schemes. In addition, we evaluate the instantaneous Mach number $M$ along the $y$-axis at $x = z = 0$, which is shown at $t = 2.5$ in Fig.~\ref{fig:tgv} (b). The results are compared against the high-resolution $2048^3$ reference data provided in~\citet{Chapelier:PF2024}. The kinetic energy plot for  5\textsuperscript{th} and 7\textsuperscript{th} order WENO and TENO schemes shows good agreement with the reference data. Lower-order schemes, such as Linear and CWENO3-Z show slight differences from the reference data. Further, looking at the Mach number $M$ distribution, we notice the best agreement with the WENO7-Z and TENO7 schemes. 



\subsection{Isentropic Vortex}\label{sec:isev}

\begin{figure*}
    \centering
    \includegraphics[width = 0.85\textwidth]{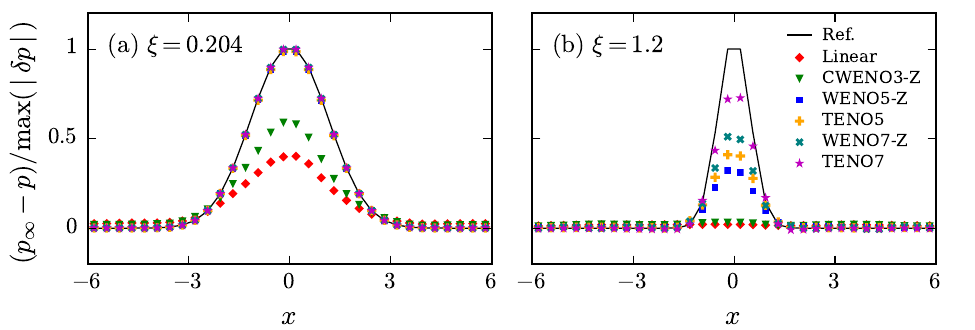}
    \caption{Normalized pressure error along the $x$-axis at $y = 0$ for the isentropic vortex test at $t = 24$. Results are shown for (a) a broad vortex with $\xi = 0.204$ and (b) a sharp vortex with $\xi = 1.2$.}
    \label{fig:isentropic}
\end{figure*}

We consider the well-known isentropic vortex problem in two dimensions~\cite{Nonomura:CF2010, Song:JCP2024}, a standard benchmark for assessing the accuracy and dissipation characteristics of numerical schemes solving the Euler equations. We perform the isentropic vortex test for $\gamma=1.4$ on a square domain $x, y \in [-6, 6]$, with periodic boundary conditions in both directions. In this setup, the vortex is convected at a constant velocity $V_0 = 0.5$, and due to the periodic domain, it returns to its original position at time $t = 24$. We consider two cases with different vortex widths: $\xi = 0.204$ and $\xi = 1.2$, while fixing the perturbation amplitude to $\epsilon = 0.3$ in both cases. The former corresponds to a broad and smooth vortex, whereas the latter results in a sharper, more compact structure that is more susceptible to numerical diffusion. We compare the performance of several high-order reconstruction schemes on a coarse $32^2$ grid. These comparisons are shown in Fig.~\ref{fig:isentropic}, where we plot the normalized pressure error along the $x$-axis at $y = 0$ for both vortex configurations. For the broad vortex ($\xi = 0.204$), the higher-order reconstructions--WENO5-Z, TENO5, WENO7-Z, and TENO7--capture the vortex well, while Linear and CWENO3-Z reconstruction introduce large diffusion. For the sharp vortex ($\xi = 1.2$), all schemes show noticeable diffusion, but TENO7 yields the least dissipative result.

\subsection{Kelvin-Helmholtz Instability}\label{sec:khi}

\begin{figure*}
    \centering
    \includegraphics[width = 0.85\textwidth]{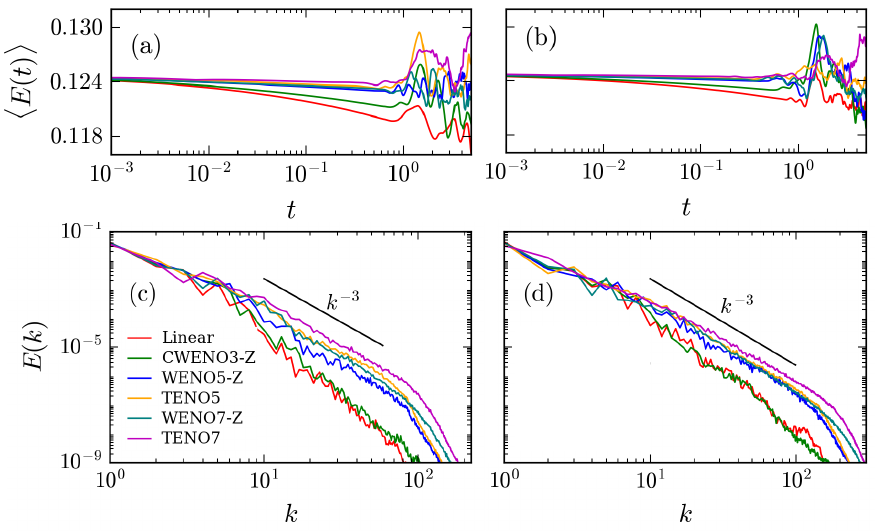}
    \caption{Kelvin-Helmholtz instability diagnostics. (a), (b): Time evolution of volume-integrated kinetic energy $E(t)$ for resolutions $512^2$ and $1024^2$, respectively. (c), (d): Velocity power spectrum at $t=5$ showing $E_k \sim k^{-3}$ scaling for higher-order reconstructions.}
    \label{fig:KH_Ekt}
\end{figure*}

We solve the two-dimensional Euler equations to study the classical Kelvin-Helmholtz (KH) instability~\cite{Zhou:book2024}. The initial condition consists of a shear layer in the horizontal velocity field and a small perturbation in the vertical velocity [$u_y = -10^{-2} \sin(4 \pi x)$] to trigger instability (see~\cite{San:CF2015}. The domain is $[-0.5, 0.5]^2$ with periodic boundary conditions in both directions. We analyze the kinetic energy $E = (u_x^2 + u_y^2)/2$ evolution and the velocity power spectrum $E(k)$ in Fig.~\ref{fig:KH_Ekt}. Subplots (a) and (b) display the time evolution of the volume-averaged kinetic energy for two resolutions: $512^2$ and $1024^2$, respectively. The TENO-based reconstructions retain more kinetic energy over time, indicating lower dissipation compared to Linear and CWENO3-Z schemes. Subplots (c) and (d) show the velocity power spectrum $E_k$ at $t=5$ for the same grid resolutions. For WENO and TENO schemes, the spectrum follows the characteristic $k^{-3}$ power law from Kraichnan–Batchelor–Leith (KBL) theory~\cite{Kraichnan:PF1967_2D,Batchelor:PF1969,Leith:JAS1971}, demonstrating accurate capturing of turbulent cascade and small-scale dynamics. In contrast, Linear and CWENO3-Z schemes show steep energy decay due to dissipative errors, failing to resolve the correct spectral behavior.




\bibliography{bib/journal, bib/book, bib/compress_journal, bib/compress_book, bib/thesis, bib/conf, bib/conf_long}

\end{document}